\documentclass[10pt,aps,prl,twocolumn,groupedaddress,showkeys]{revtex4-1}
\usepackage{amsmath}
\usepackage{graphicx}
\DeclareGraphicsExtensions{.pdf,.eps,.png,.jpg,.mps}
\usepackage{epstopdf}
\usepackage{threeparttable}
\usepackage{color}

\begin{document}

\title{Waveguide-integrated, plasmonic enhanced graphene photodetectors}
\author{J. E. Muench$^1$, A. Ruocco$^1$, M. A. Giambra$^2$, V. Miseikis$^{2,3,4}$, D. Zhang$^1$, J. Wang$^1$, H. F. Y. Watson$^1$, G. C. Park$^1$, S. Akhavan$^1$, V. Sorianello$^2$, M. Midrio$^5$, A. Tomadin$^6$, C. Coletti$^{3,4}$, M. Romagnoli$^2$, A. C. Ferrari$^1$}
\email[E-mail: ]{acf26@eng.cam.ac.uk}
\author{I. Goykhman$^7$}
\affiliation{$^1$Cambridge Graphene Centre, University of Cambridge, Cambridge CB3 0FA, UK}
\affiliation{$^2$Consorzio Nazionale per le Telecomunicazioni, Photonic Networks and Technologies National Laboratory, 56124 Pisa, Italy}
\affiliation{$^3$Center for Nanotechnology Innovation $@$ NEST, Istituto Italiano di Tecnologia, 56127 Pisa, Italy}
\affiliation{$^4$Graphene Labs, Istituto Italiano di Tecnologia, 16163 Genova, Italy}
\affiliation{$^5$Consorzio Nazionale per le Telecomunicazioni, University of Udine, 33100 Udine, Italy}
\affiliation{$^6$Dipartimento di Fisica, Universit\`a di Pisa, Largo Bruno Pontecorvo 3, 56127 Pisa, Italy}
\affiliation{$^7$Micro- and Nanoelectronics Research Center, Technion, Haifa 320000, Israel}

\begin{abstract}
We present a micrometer scale, on-chip integrated, plasmonic enhanced graphene photodetector (GPD) for telecom wavelengths operating at zero dark current. The GPD is designed and optimized to directly generate a photovoltage and has an external responsivity$\sim$12.2V/W with a 3dB bandwidth$\sim $42GHz. We utilize Au split-gates with a$\sim$100nm gap to electrostatically create a p-n-junction and simultaneously guide a surface plasmon polariton gap-mode. This increases light-graphene interaction and optical absorption and results in an increased electronic temperature and steeper temperature gradient across the GPD channel. This paves the way to compact, on-chip integrated, power-efficient graphene based photodetectors for receivers in tele and datacom modules.
\end{abstract}
\maketitle
The ever-growing demand for global data traffic\cite{cisco2017visual} is driving the development of next generation communication standards\cite{Andrews2014What,Osseiran2014Scenarios}. The increasing numbers of connected devices\cite{Arm2017route}, the need for new functionalities, and the development of high-performance computing\cite{Rumley2015Silicon,zhou2018development} require optical communication systems performing at higher speeds, with improved energy-efficiency, whilst maintaining scalability and cost-effective manufacturing. Si photonics\cite{reed2008silicon,thomson2016roadmap,absil2015silicon} offers the prospect of dense (nanoscale) integration\cite{Atabaki2018Integrating} relying on mature, low-cost (based on complementary metal-oxide-semiconductor (CMOS) fabrication processes) manufacturing\cite{thomson2016roadmap,absil2015silicon}, making it one of the key technologies for short-reach ($<$10km) optical interconnects\cite{biberman2012optical} beyond currently employed lithium niobate\cite{Wooten2000review} and indium phosphate\cite{Nagarajan2010InP}.

A variety of functionalities have been developed and demonstrated in Si photonics for local optical interconnects\cite{biberman2012optical}. Electro-optic modulators based on carrier-depletion (phase-modulation) in Si\cite{Reed2010silicon,reed2014recent} or the Franz-Keldysh effect\cite{seraphin1965franz} (amplitude-modulation) in strained Si-Ge\cite{liu2008waveguide,srinivasan201656} encode information into optical signals at telecom wavelengths ($\lambda=$1.3-1.6$\mu$m). On the receiver side, Ge\cite{Michel2010High} or bonded III-V\cite{Hawkins1997High,chang2010integrated} photodetectors (PD) are needed for optical-to-electrical signal conversion, since the telecom photon energies are not sufficient for direct (band-to-band) photodetection in Si\cite{chrostowski2015silicon}.

On-chip integrated Ge PDs\cite{vivien2009Ghz,derose2011ultra,vivien2012zero,novack2013germanium,Chen2016Bias} are standard components in Si photonics foundries\cite{thomson2016roadmap,absil2015silicon,chrostowski2015silicon}. Their external responsivities (in A/W), $R_I=I_{\text{ph}}$/$P_{\text{in}}$, where $I_{\text{ph}}$ is the photocurrent and $P_{\text{in}}$ is the incident optical power, can exceed 1A/W\cite{vivien2009Ghz,thomson2016roadmap} and their bandwidth can reach 60GHz\cite{Chen2016Bias,vivien2012zero,novack2013germanium}. Following the development of high temperature ($>600^\circ$C)\cite{Michel2010High} heterogeneous integration of Ge-on-Si using epitaxial growth and cyclic thermal annealing\cite{Michel2010High,wang2011ge,ye2014germanium}, the concentration of defects and threading dislocations in Ge epilayers and at Si/Ge interfaces can be reduced\cite{Michel2010High}, resulting in low ($<$10nA\cite{Chen2016Bias,absil2015silicon}) dark current in waveguide integrated Ge p-i-n photodiodes\cite{derose2011ultra,Chen2016Bias}. However, Ge-on-Si integration is a complex process\cite{Michel2010High,ye2014germanium,chrostowski2015silicon}, as the lattice mismatch between Si and Ge\cite{Michel2010High}, ion implantation\cite{vivien2009Ghz,vivien2012zero}, thermal budget (i.e. thermal energy transfer to the wafer) management\cite{chrostowski2015silicon}, and the non-planarity of Ge layers\cite{ye2014germanium} require dedicated solutions during device fabrication\cite{absil2015silicon}. The charge carrier mobility $\mu$ in Si and the dislocations and defects in grown\cite{Michel2010High} or evaporated\cite{sorianello2012high} Ge layers set intrinsic limitations that prevent further improvements to the operation speed of Ge PDs without compromising $R_I$\cite{novack2013germanium,absil2015silicon}. These shortcomings, together with the spectrally limited operation regime (band edge in Ge$\sim 1.57\mu$m\cite{chrostowski2015silicon}, which can be extended to$\sim 1.62\mu$m\cite{liu2005tensile} at the expense of $R_I$), and the incompatibility of Ge epitaxy for monolithic integration with other material platforms, such as SiN, are amongst the main limitations for Ge PDs\cite{thomson2016roadmap}. Thus, novel solutions for PDs, integrated with Si photonics, at telecom bands are needed.

Graphene is a promising candidate for on-chip integrated photonics\cite{Romagnoli2018Graphene,Liu2011graphene,Liu2012double,Hu2016broadband,phare2015graphene,sorianello2015design,Sorianello2018graphene,Sun2016optical,Gan2013chip,Pospischil2013CMOS,Wang2013high,Goykhman2016On,Schall2014GBits,Schuler2016Controlled,schuler2018graphene,Shiue2015High,Schall2017graphene,Schall2018record,Ding2018ultra,Ma2018Plasmonically,ma2018compact}.
The advantages of single-layer graphene (SLG) for photonics stem from its superior optoelectronic properties\cite{Bonaccorso2010Graphene}. These include high-speed ($>$200GHz\cite{urich2011intrinsic}) operation\cite{Xia2009ultrafast}, broadband (ultraviolet to far-infrared) absorption\cite{Nair2008Fine,dawlaty2008measurement,Koppens2014Photodetectors}, efficient optical modulation (electro-optical index change $\Delta n_{\text{eff}}>10^{-3}$)\cite{Romagnoli2018Graphene,Liu2011graphene,Liu2012double,Hu2016broadband,phare2015graphene,sorianello2015design,Sorianello2018graphene}, CMOS compatibility\cite{goossens2017broadband,Pospischil2013CMOS} and integrability\cite{Romagnoli2018Graphene,Youngblood2016integration,Liu2013silicon} with different on-chip photonics platforms, such as silicon-on-insulator (SOI)\cite{Liu2011graphene} and SiN\cite{phare2015graphene}. In the case of waveguide-integrated graphene PDs (GPDs)\cite{Gan2013chip,Pospischil2013CMOS,Wang2013high,Goykhman2016On,Schall2014GBits,Schuler2016Controlled,schuler2018graphene,Shiue2015High,Schall2017graphene,Schall2018record,Ding2018ultra,Ma2018Plasmonically,ma2018compact}, high speeds up to 128GHz\cite{Schall2018record}, wafer-scale integration\cite{Schall2017graphene} and $R_I\sim$0.4-0.5A/W\cite{Shiue2015High,Goykhman2016On,Ding2018ultra,Ma2018Plasmonically} were reported. GPDs can offer broadband detection across multiple telecommunication channels (O-band$\sim$1.31$\mu$m to U-band$\sim$1.65$\mu$m)\cite{Pospischil2013CMOS}, bias-free operation\cite{tielrooij2015generation}, and direct generation of photovoltage\cite{tielrooij2015generation,Schuler2016Controlled}. The latter opens up the possibility of building GPDs without the noise contribution of dark current\cite{schuler2018graphene,Romagnoli2018Graphene} and eliminates the need of noise-prone trans-impedance amplifier (TIA) to convert current-to-voltage in the read-out electronics\cite{Romagnoli2018Graphene}.

GPDs can be built exploiting different mechanisms: photo-voltaic (PV)\cite{Xia2009ultrafast,Mueller2010graphene,Echtermeyer2014Photothermoelectric}, photo-thermoelectric (PTE)\cite{Song2011Hot,gabor2011hot,Echtermeyer2014Photothermoelectric}, photo-gating\cite{Konstantatos2012hybrid}, plasma-wave assisted\cite{Vicarelli2012graphene} and photo-bolometric (PB)\cite{freitag2013photoconductivity,Sassi2016graphene}. The dominating effect for a given GPD depends on device configuration, design geometry, and way of operation\cite{Huo2018Recent,Echtermeyer2014Photothermoelectric}. For telecom applications, where high-speed (tens GHz) operation is one of the key requirements\cite{thomson2016roadmap,Romagnoli2018Graphene}, PV, PTE and PB are typically considered for waveguide-integrated GPDs\cite{Romagnoli2018Graphene}, taking advantage of the ultra-fast ($\sim$fs-ps) carrier dynamics in SLG\cite{Brida2013ultrafast,tomadin2013nonequilibrium}.

PTE is ideal for PD operation in a voltage mode. In optically illuminated SLG, electron-electron scattering drives the formation of a 'hot' (optically excited)-carrier distribution, described by the Fermi-Dirac function\cite{kittel1996introduction}, within $<$50fs\cite{Brida2013ultrafast}. This can remain at elevated temperatures $T_e$, well above the lattice temperature $T_l$, over$\sim$2-4ps  time scales\cite{Brida2013ultrafast}, before reaching thermal equilibrium via phonons interaction\cite{tomadin2013nonequilibrium,bonini2007phonon,lazzeri2005electron}. In this hyperthermal state, a photovoltage $V_\text{ph}$ is generated by a thermo-electric current as for the Seebeck effect\cite{gabor2011hot}, if both a temperature and chemical potential gradient are present in the SLG channel. The sign and magnitude of $V_\text{ph}$ depend on the Seebeck coefficient ($S$), i.e. the proportionality constant between temperature change and induced photovoltage\cite{ashcroft1976solid}, and $T_\text{e}$ profile in the SLG channel\cite{gabor2011hot}:
\begin{equation} \label{eq:VPTE}
V_\text{ph}=\int S(x) \cdot \nabla T_e(x) \; \mathrm{d} x
\end{equation}
where $x$ is the coordinate along the channel from drain to source, and $S$ is given by Mott's formula\cite{ashcroft1976solid,gabor2011hot, Song2011Hot,Echtermeyer2014Photothermoelectric}:
\begin{equation} \label{eq:MottFormula}
S(x) = - \frac{\pi^2k_\text{B}^2T_e}{3e}\frac{1}{\sigma(x)}\frac{\mathrm{d}\sigma(x)}{\mathrm{d}\mu_c}
\end{equation}
with $\sigma(x)$ the conductivity, $k_\text{B}$ the Boltzmann constant, $e$ the electron charge and $\mu_c$ the chemical potential ($\mu_c=E_{F}$ at $T_{e}=0$\cite{kittel1996introduction}, with $E_{F}$ the Fermi energy).

PTE-GPDs have been reported in vertically-illuminated\cite{gabor2011hot,Ma2014Competing,freitag2013increased,Herring2014photoresponse,castilla2019fast} and waveguide-integrated\cite{Schuler2016Controlled,schuler2018graphene,Shiue2015High}configurations. The latter used SLG flakes prepared by micromechanical cleavage (MC) of graphite\cite{Novoselov2005Two}, with typical device length of tens of $\mu$m\cite{Schuler2016Controlled,schuler2018graphene,Shiue2015High}. They have external voltage responsivities, defined as $R_V=V_\text{ph}/P_\text{in}$, up to$\sim$4.7V/W\cite{schuler2018graphene} (at zero bias) with speeds up to 65GHz\cite{Schuler2016Controlled}. Depending on PTE-GPD design configuration and the requirements of the read-out electronics (i.e. output photo-signal to be measured as current or voltage), the responsivity can be characterized in terms of $R_I$ or $R_V$. The photovoltage generated by the Seebeck effect is associated with a thermoelectric current across the PD by a Ohmic relation\cite{tielrooij2015generation,freitag2013increased,Schuler2016Controlled,schuler2018graphene} $I_\text{ph}=V_\text{ph}/R$, with $R$ the resistance.

To increase $R_V$ for PTE-GPDs, Eq.\ref{eq:VPTE} suggests two strategies: 1) maximize $S$; 2) maximize the $T_\text{e}$ gradient profile in the SLG channel. The former is related to $\mu$ via Eq.\ref{eq:MottFormula} and the Drude conductivity\cite{ashcroft1976solid}, $\sigma=e\mu n$, where $n$ is the charge carrier concentration. Thus, $S$ can be improved by using high-mobility SLG, e.g. encapsulating SLG in hBN\cite{Wang2013One,purdie2018cleaning,DeFazio2019high}, using single-crystals\cite{Miseikis2017deterministic,DeFazio2019high}, or large (tens $\mu$m) domain-size\cite{li2010graphene}, in combination with a transfer processes that avoid contamination\cite{purdie2018cleaning,Wang2016Support}, strain\cite{Wang2016Support}, and cracks\cite{suk2011transfer}. Ref.\cite{Romagnoli2018Graphene} suggested that $\mu>10^4$cm$^2$V$^{-1}$s$^{-1}$ could enable $R_V>100$V/W. The $T_e$ gradient can be increased by creating a spatially confined, localized, heat source\cite{Schuler2016Controlled} generated by enhanced optical absorption in SLG over compact ($<10\mu$m) device lengths\cite{Ding2018ultra,Ma2018Plasmonically}. This could be achieved by integrating plasmonic nanostructures\cite{Goykhman2011Locally,Goykhman2012Waveguide,Salamin2018GHz,Echtermeyer2011Strong,fang2012graphene,Echtermeyer2016Surface}. Sub-wavelength plasmonic confinement and associated enhancement of near-field light-matter interaction were used to boost $R_I$ in Si-plasmonic PDs\cite{Goykhman2011Locally,Goykhman2012Waveguide}, plasmonic-Ge PDs\cite{Salamin2018GHz}, plasmonic decorated GPDs for free-space\cite{Echtermeyer2011Strong,fang2012graphene,Echtermeyer2016Surface} and waveguide-integrated\cite{Goykhman2016On,chen2017three,Ding2018ultra,Ma2018Plasmonically,ma2018compact} configurations. Refs.\cite{Ding2018ultra,Ma2018Plasmonically,ma2018compact} reported plasmonic enhanced on-chip GPDs based on PV\cite{Ding2018ultra,ma2018compact} and PB\cite{Ma2018Plasmonically,ma2018compact} with $R_I\sim0.5$A/W and bandwidth$\sim110$GHz at 1.55$\mu$m for source-drain bias$<$1V.
\begin{figure*}
\centerline{\includegraphics[width=180mm]{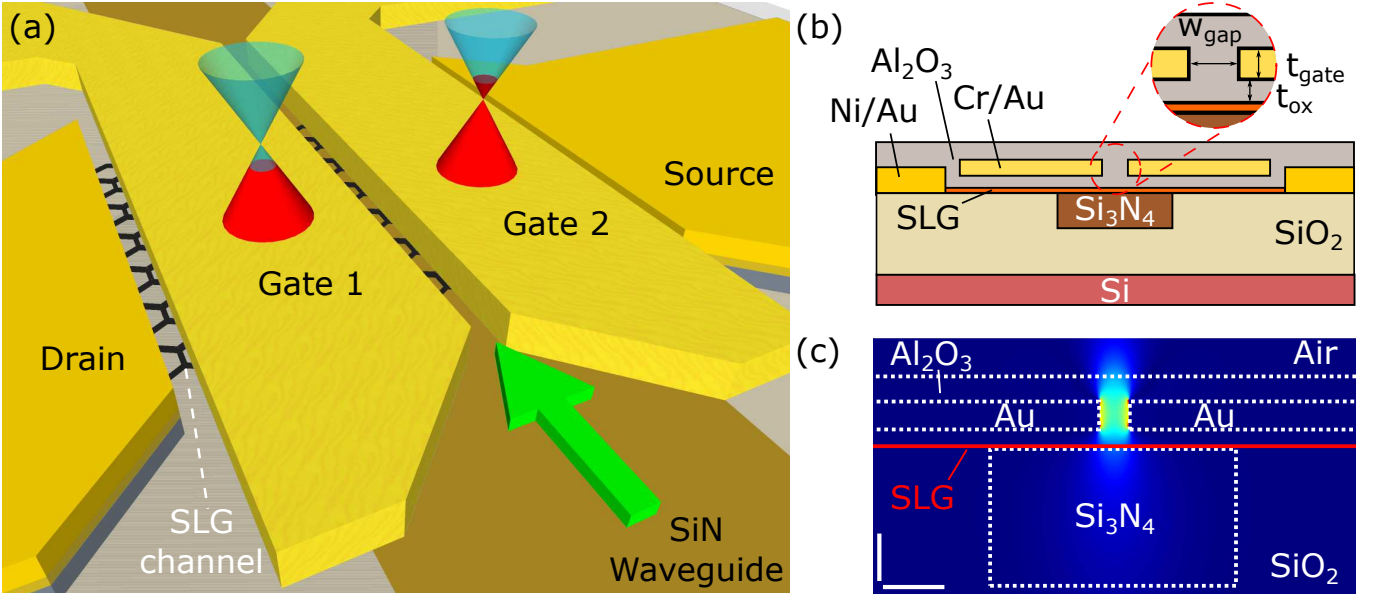}}
\caption{a) Schematic view of our GPD: SLG on SiN waveguide (brown) with split-gates, acting as plasmonic slot waveguide, to create a p-n junction in the channel (as depicted by the Dirac cones above the gates). The green arrow indicates the light propagation direction. b) Schematic cross-section of the GPD active region. c) Simulated electric field (E$_x$, in-plane) distribution of the fundamental SPP waveguide mode. For clarity, only the field component parallel to the SLG channel is shown. The vertical and horizontal scale bars are 100 and 250nm}
\label{Concept}
\end{figure*}

Here, we report compact ($\sim$0.5-4$\mu$m), PTE-based, waveguide-integrated, plasmonic-enhanced GPDs for telecom wavelengths with $R_V\sim12.2$V/W at zero source-drain bias and zero dark current, with a 3dB cutoff frequency$\sim42$GHz. To the best of our knowledge, this is the largest $R_V$ to date for waveguide-integrated GPDs operating in voltage mode. We use SLG grown by chemical vapor deposition (CVD) and transferred onto low-loss ($\sim1$dB/cm) planarized (i.e. fully-embedded in polished cladding\cite{reed2008silicon}) SiN waveguides with a semi-dry (i.e. combining wet de-lamination from the growth substrate with dry lamination onto the target substrate) transfer\cite{Miseikis2017deterministic}, unlike previous PTE GPDs exploiting non-scalable MC SLG\cite{Schuler2016Controlled,schuler2018graphene}. Our design relies on Au split-gates to electrostatically create a p-n junction in the SLG channel, as well as to guide a confined SPP waveguide mode. By leveraging optical field enhancement and plasmonic confinement in the gap, we increase light-SLG interaction and optical absorption in the p-n junction region, resulting in a confined electrons heat source, compact device length, and increased $R_V$. This combines high-performance (large $R_V$, high-speed, bias-free, compact, direct $V_\text{ph}$ read-out) PTE GPDs in the telecom range with scalable fabrication, paving the way for graphene integrated receivers for next-generation transceivers.

The design of our GPD is schematically shown in Fig.\ref{Concept}a,b. It comprises a SLG channel on a SiN waveguide supporting a transverse-electric (TE, in-plane) polarized fundamental waveguide mode. Two Au gates are placed above the channel, separated from the SLG by an Al$_2$O$_3$ dielectric spacer and centrally aligned with respect to the waveguide. When this split-gate structure is DC (direct current) biased, it forms a p-n junction, Fig.\ref{Concept}a, and creates a $S$ profile in the SLG channel, as for Eq.\ref{eq:MottFormula}. When an on-chip guided light signal reaches the PD area, it is evanescently coupled from the SiN waveguide to the split-gate, which acts as SPP waveguide, Fig.\ref{Concept}c. The plasmonic guiding with light confinement in the gap (width $w_\text{gap}$ $\sim$100nm) leads to enhanced optical absorption and a localized hot electron distribution with a $T_\text{e}$ gradient in the p-n junction (gap) region. The coupling efficiency, $P_{\text{out}}/P_{\text{in}}$, where $P_{\text{out}}$ is the power transferred between two optical components, from photonic to plasmonic waveguide mode can be optimized by tailoring $w_\text{gap}$ and dielectric spacer thickness ($t_\text{ox}$).
\begin{figure*}
\centerline{\includegraphics[width=180mm]{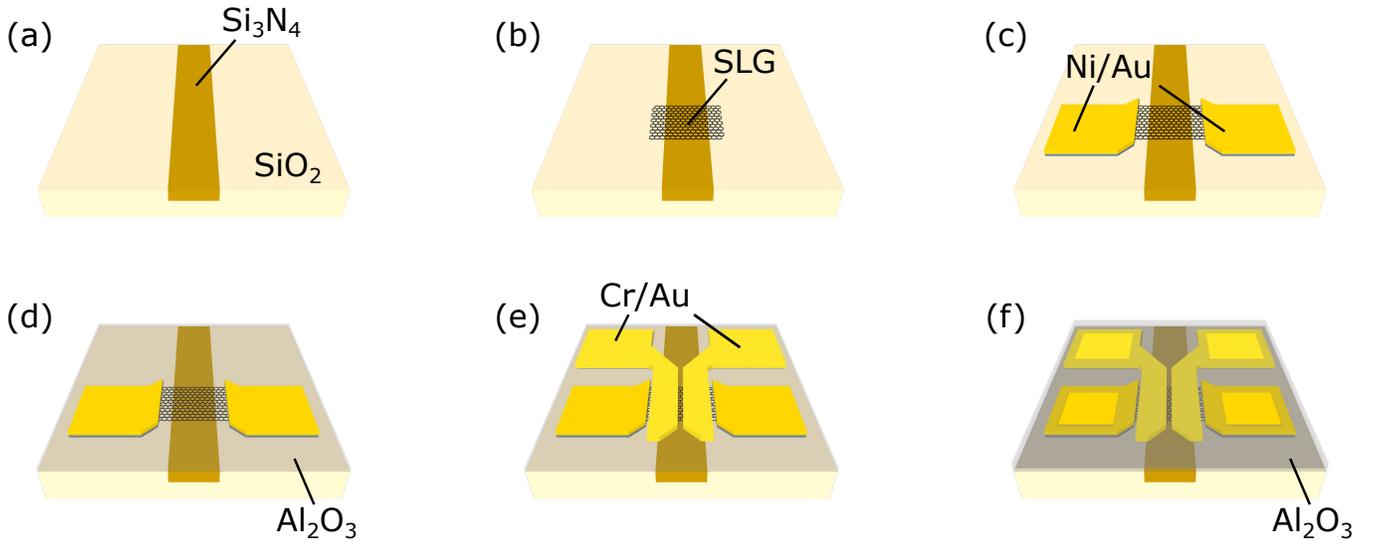}}
\caption{a) Planarized SiN waveguide. b) SLG transfer and shaping by oxygen plasma etch. c) Ni/Au contacts to SLG channel. d) $\text{Al}_{2}\text{O}_{3}$ gate oxide deposition. e) Cr/Au evaporation of split-gate structure. f) $\text{Al}_{2}\text{O}_{3}$ encapsulation and contact pads opening}
\label{FabricationProcess}
\end{figure*}
\begin{figure*}
\centerline{\includegraphics[width=180mm]{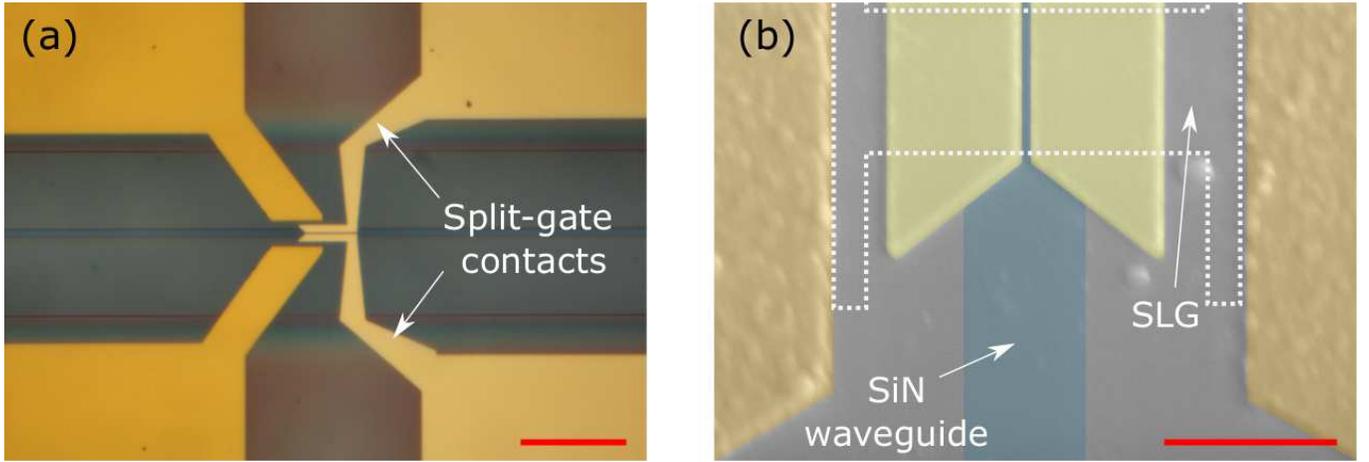}}
\caption{a) Optical image of a GPD. Scale bar: 20$\mu$m. b) Scanning electron micrograph of split-gate. False colors: brown, Ni/Au contacts; yellow, Cr/Au gates; green, planarised SiN waveguide; white dashed line, SLG channel. Scale bar: 2$\mu$m}
\label{SEM}
\end{figure*}

To optimize the cross section parameters at $\lambda$=1.55$\mu$m, we perform optical simulations using a commercial finite difference solver tool (Lumerical MODE). After selecting the fundamental gap plasmon mode for a given design and $\lambda$, we extract the optical electric field distribution in the SLG channel to model the absorbed power density that generates the hot carrier distribution as time-averaged electric power dissipation density\cite{desiatov2014direct,shin2012instantaneous}, which we refer to as Joule heat source ($J$) hereafter. After normalization to an input power of 1$\mu$W, this is used in the heat equation\cite{Song2011Hot,Ma2014Competing,Shiue2015High}:
\begin{equation} \label{eq:heatequation}
-\kappa_e(x) \left[\frac{\mathrm{d}^2}{\mathrm{d}x^2} \Delta T_e(x) - \frac{1}{\xi^2} \Delta T_e(x) \right]=J\left(x\right)
\end{equation}
where $\Delta T_e(x)=T_e(x)-T_l$ is the local temperature fluctuation, $\xi$ is the cooling length (see Methods) and $\kappa_e(x)$ is the electronic thermal conductivity (see Methods). Eq.\ref{eq:heatequation} gives the $T_e(x)$ profile along the SLG channel. The $S(x)$ profile from Eq.\ref{eq:MottFormula} is used in Eq.\ref{eq:VPTE} to obtain $V_\text{ph}$. The device parameters are chosen to maximize $V_\text{ph}$.

A second aspect of device design concerns the coupling between the dielectric and plasmonic waveguides, as well as the positioning and width of the SLG channel along the split-gate. Taper-assisted butt-coupling (end-to-end alignment) was reported to yield the lowest insertion loss ($<0.6$dB)\cite{tian2009broadband} for the transition from optical to plasmonic modes. However, since evanescent coupling (lateral or vertical alignment) provides simpler fabrication\cite{dabos2018water} and greater flexibility for the placement of devices on top of integrated optical circuits\cite{li2010structurally}, we use this here, Fig.\ref{Concept}a-c. To obtain the largest $R_V$, the electric field distribution along the propagation directions needs to be considered. Light absorption in SLG or in the plasmonic structure along the device leads to an exponential decay of optical power\cite{Youngblood2016integration}. Thus, the increase in $T_e$ follows the same decay. The resulting photovoltage drop at different points along the device results in an averaged potential difference between source and drain contacts. To optimize $R_V$, a compact ($<$10$\mu$m) device with optimized peak absorption and minimal drop-off is preferable. We thus perform finite-difference time domain (FDTD) simulations in Lumerical FDTD (see Methods). The co-existence of plasmonic and dielectric waveguide leads to oscillating power exchange between both structures\cite{delacour2010efficient,li2010structurally,dabos2018water}. For the highest coupling efficiency, the vertical distance between these waveguides is typically$>150$nm\cite{li2010structurally,dabos2018water}, exploiting interference between quasi-even and quasi-odd eigenmodes\cite{delacour2010efficient,li2010structurally,dabos2018water}. In our design, we keep this separation small (tens nm) to ensure overlap between SLG and gap plasmon mode, to avoid a long ($>5\mu$m) coupling length\cite{dabos2018water}, and to create a sharper concentration of power close to the front of the plasmonic structure. The SLG channel is placed accordingly, after a short ($\sim$1$\mu$m) taper at the front of the SPP waveguide to reduce mode mismatch.
\begin{figure}
\centerline{\includegraphics[width=90mm]{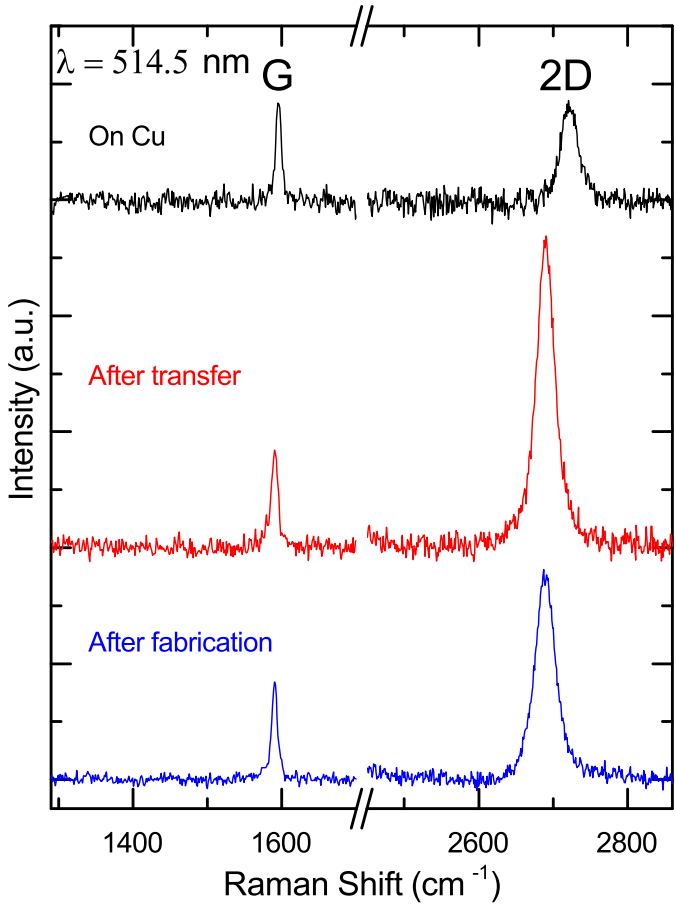}}
\caption{Raman spectra at 514.5 nm for SLG as grown on Cu (black), after transfer onto the SiN waveguide (red), and after device fabrication (blue). All spectra are normalised to the intensity of the G peak, $I(\text{G})$, and are shown after subtraction of the substrate signals.}
\label{Raman}
\end{figure}
\begin{figure*}
\centerline{\includegraphics[width=170mm]{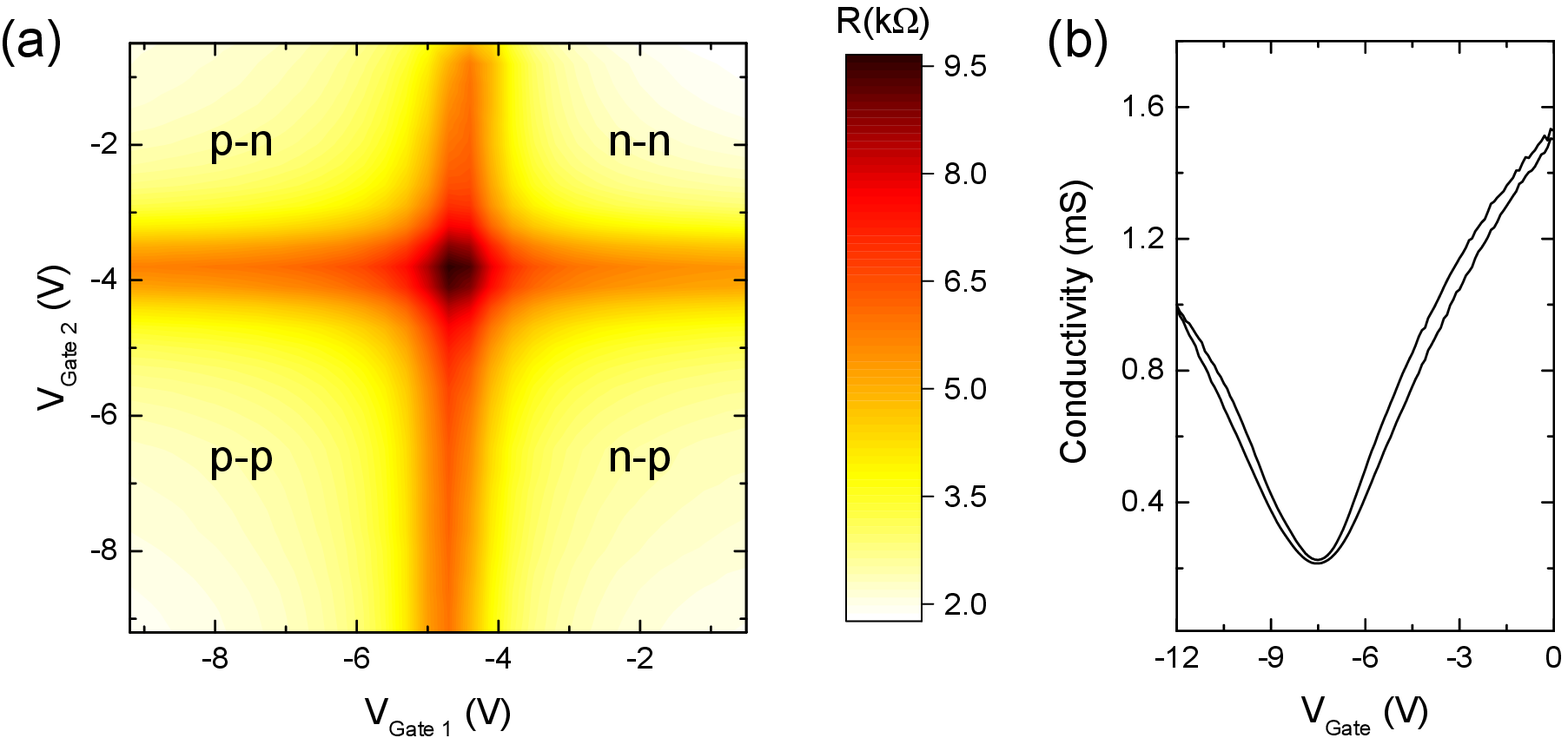}}
\caption{(a) GPD channel resistance as function of split-gate voltages. (b) Conductivity vs. gate voltage from a 4-terminal measurement on test Hall bars}
\label{Electrical}
\end{figure*}

Fig.\ref{FabricationProcess} summarizes the fabrication process of our GPDs. Planarized SiN waveguides, Fig.\ref{FabricationProcess}a, (260nm high, width 0.8-1.5$\mu$m) on 15$\mu$m $\text{SiO}_2$ are fabricated as follows. The SiN layer is first deposited by low-pressure (LP) CVD. The SiN photonic waveguides are then defined by electron beam lithography (EBL) and reactive ion etching. For surface planarization, a 1.6 $\mu$m thick boron-phosphorus tetraethyl orthosilicate (BPTEOS) layer is deposited as top cladding and subsequently etched to a final thickness$\sim 20$nm on top of the SiN waveguides, avoiding chemical mechanical polishing. SLG is grown on pre-patterned, electropolished Cu with Cr nucleation sites as for Ref.\cite{Miseikis2017deterministic}. After an initial annealing in argon (10mins), SLG growth is initiated at 25mbar with argon, hydrogen, and methane flowing at 900, 100, and 1 standard cubic centimeters per minute (sccm), respectively. After growth, SLG single crystals are placed onto the photonic chips by semi-dry transfer\cite{Miseikis2017deterministic}, comprising the spin-coating of a poly(methyl methacrylate) (PMMA) support layer, the attachment of a Kapton frame for handling, electrochemical delamination of SLG in sodium hydroxide, and the lamination onto the target substrate with the help of a micro-manipulator to align the crystals with the photonic structures. A PMMA etch mask is then used to shape the device channel and remove excess SLG over grating coupler and waveguides, defined using EBL. This is followed by oxygen plasma etching at 3W, Fig.\ref{FabricationProcess}b. Next, contacts are defined by another EBL step. Metallization (15nm Ni/40nm Au) is done by sputtering, thermal evaporation and lift-off in acetone, Fig.\ref{FabricationProcess}c. 30nm $\text{Al}_2\text{O}_3$ is used as gate oxide, via atomic layer deposition (ALD), Fig.\ref{FabricationProcess}d. An additional EBL step and electron beam evaporation are used to fabricate the plasmonic split gates, Fig.\ref{FabricationProcess}e. To encapsulate the device and prevent air breakdown in the gap between gate contacts when$\sim$10V is applied, we deposit another 40nm $\text{Al}_2\text{O}_3$ by ALD. A laser writer is used to define an etch mask to open access to all contacts, Fig.\ref{FabricationProcess}f.

The quality of SLG is monitored by Raman spectroscopy at all critical points during the fabrication process, using a Renishaw InVia equipped with a 50x objective (numerical aperture NA=0.75) at 514.5nm with power below 0.5mW to exclude heating effects and risk of damage. Representative spectra of SLG on Cu (after removal of Cu background photoluminescence\cite{lagatsky20132}), after transfer onto the waveguide, and after complete device fabrication, are shown in Fig.\ref{Raman}. The absence of a D peak confirms negligible defects are introduced during fabrication. The 2D peaks are single-Lorentzian, confirming the presence of SLG\cite{Ferrari2006Raman,Ferrari2013Raman}. On Cu, the position and full width at half-maximum of the G peak are Pos(G)$\sim$1595cm$^{-1}$ and FWHM(G)$\sim8$cm$^{-1}$. The position of the 2D peak, Pos(2D), is$\sim$2721cm$^{-1}$ with FWHM(2D)$\sim$27cm$^{-1}$. The 2D to G peak intensity, $I(\text{2D})/I(\text{G})$, and area, $A(\text{2D})/A(\text{G})$, ratios are$\sim1$ and$\sim3.2$. After transfer, Pos(G)$\sim1590$cm$^{-1}$, FWHM(G)$\sim$10cm$^{-1}$, Pos(2D)$\sim2690$cm$^{-1}$, FWHM(2D)$\sim$28cm$^{-1}$, $I(\text{2D})/I(\text{G})\sim3.2$ and $A(\text{2D})/A(\text{G})\sim8.6$. This corresponds to$\sim250$meV doping\cite{Das2008Monitoring,Basko2009Electron} and a carrier concentration$\sim$4$\times 10^{12}$cm$^{-2}$. After the final encapsulation, Pos(G)$\sim$1590cm$^{-1}$, FWHM(G)$\sim$9cm$^{-1}$, Pos(2D)$\sim$2689cm$^{-1}$, FWHM(2D)$\sim$30cm$^{-1}$, $I(\text{2D})/I(\text{G})\sim2.2$ and $A(\text{2D})/A(\text{G})\sim7.6$, indicating$\sim$350meV ($n\sim$7$\times10^{12}$cm$^{-2}$) doping.
\begin{figure*}
\centerline{\includegraphics[width=180mm]{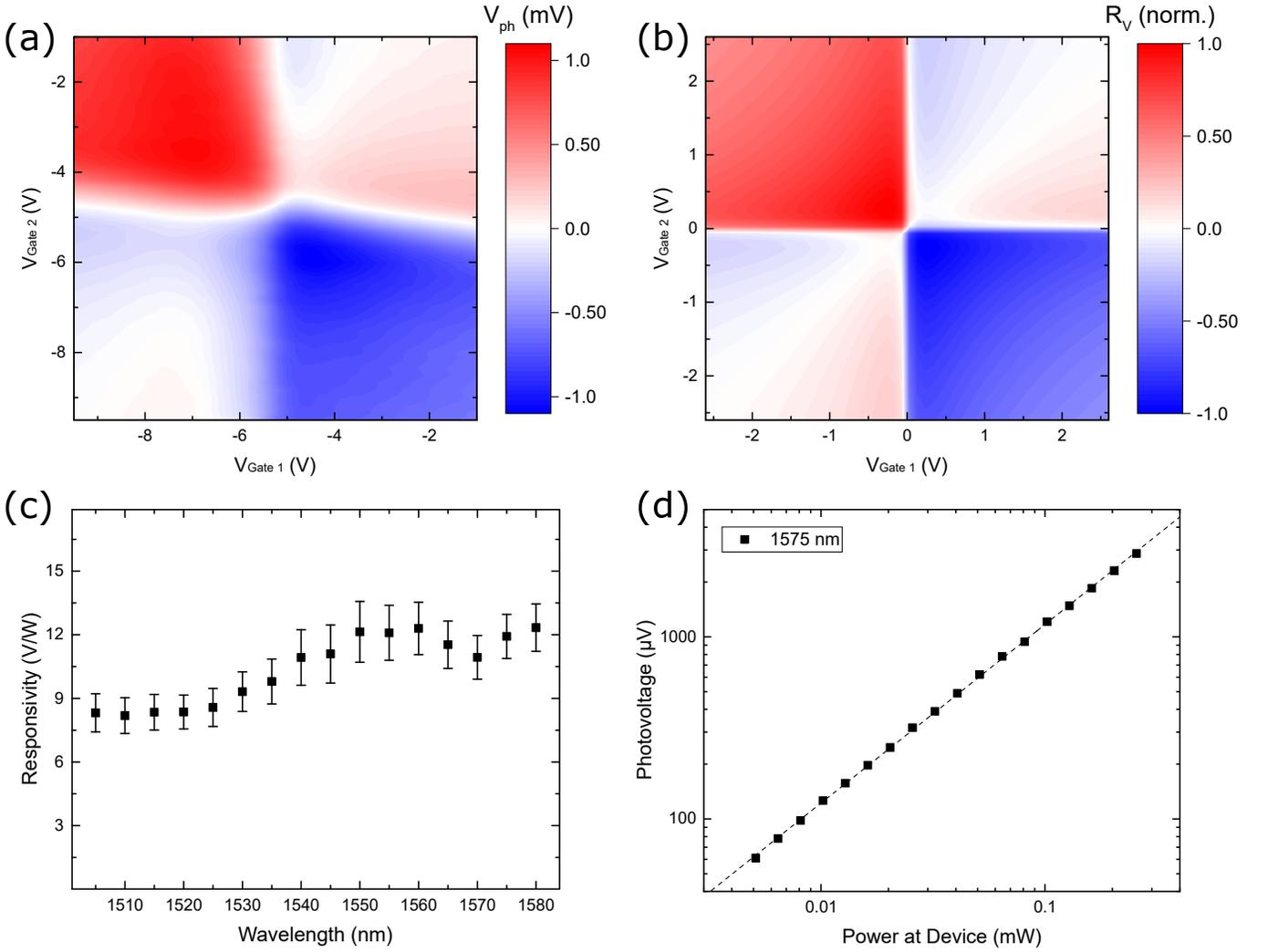}}
\caption{a) Experimental photovoltage map for zero bias. b) Simulated responsivity. c) Wavelength dependence of responsivity. d) Power dependence of responsivity at 1575 nm.}
\label{OpticalSgl}
\end{figure*}
\begin{figure*}
\centerline{\includegraphics[width=180mm]{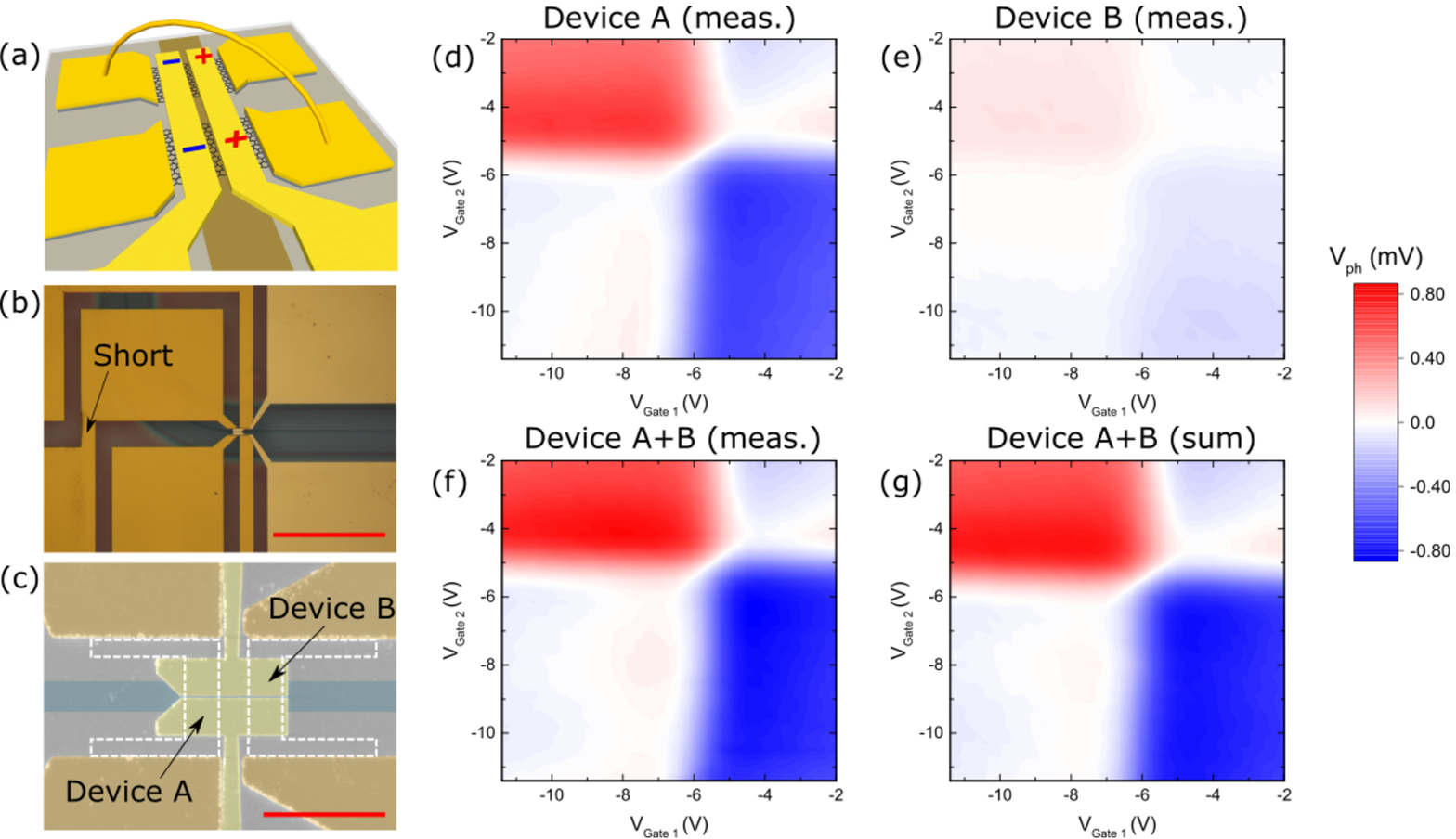}}
\caption{a) Schematic of test structure to measure two GPDs in series. b) Optical image of contact pad short to connect both GPDs in series. Scale bar: 80$\mu$m. c) SEM image of active region of test structure. False colors: brown, Ni/Au contacts; yellow, Cr/Au split gate; green, planarized SiN waveguide. The SLG channels are indicated by white dashed lines. Scale bar: 5$\mu$m. d) Photovoltage map of device A at the start of the SPP waveguide. e) Photovoltage map of device B at the end of the SPP waveguide. f) Photovoltage map of devices A and B in series. g) Sum of the photovoltage maps for devices A and B}
\label{OpticalDualConcept}
\end{figure*}

To determine the DC operating point, we perform electrical characterizations by sweeping the split-gate voltages ($V_\text{Gate 1}$, $V_\text{Gate 2}$) while measuring the device current $I_{DS}$ under a constant source-drain bias $V_{DS}$=1mV, using DC probes on micromanipulators and two source measure units. To record the static photoresponse, we add two fibre probes and couple continuous wave (CW) transverse-electric (TE) polarized light at 1.50-1.58$\mu$m from a tunable laser into the SiN waveguide via an optical fibre and a grating coupler (GC). While $V_\text{ph}$ is recorded across the unbiased ($V_{DS}$=0V) channel as a function of $V_\text{Gate 1}$ and $V_\text{Gate 2}$, using a lock-in amplifier under internal modulation (square wave, ON-OFF) of the laser with 200Hz, we monitor the transmission with a second fibre positioned over the output GC and connected to an external InGaAs power meter to ensure constant $P_\text{in}$.

Fig.\ref{Electrical}a plots the $R$ map of a typical device as a function of $V_\text{Gate 1}$, $V_\text{Gate 2}$. This shows a four-fold pattern, corresponding to the four doping constellations (p-n, n-p, n-n, p-p) in the SLG channel for different combinations of gate voltages. The map is symmetric with a maximum $R\sim$9k$\Omega$ at the crossing of the charge neutrality point (CNP), between -4 and -5V. This corresponds to n-doping of the unbiased SLG channel with n$\sim$7$\times 10^{12}$cm$^{-2}$ ($\sim$350meV). $R$ has contributions from channel ($R_{ch}$) and contact ($R_{c}$) resistances. $R_{ch}$ includes a fixed contribution from ungated SLG regions and a gate-dependent contribution from channel segments underneath the split-gates. The gate-dependent variation in $R$ in Fig.\ref{Electrical}a suggests $R_{ch}$ as the dominant factor. This is consistent with our contact resistivity ($<$1k$\Omega\mu$m) for CVD SLG and the calculated $R$ based on channel geometry and sheet resistance obtained from independent four-terminal measurements on reference Hall bars. From these we also extract an average $\mu\sim$2000cm$^2$V$^{-1}$s$^{-1}$ from linear fits of the conductivity via\cite{Novoselov2004Electric} $\mu=\vert \mathrm{d}\sigma/\mathrm{d}\text{V}_\text{Gate}\vert / C_\text{ox}$, where $C_\text{ox}$ is the top gate capacitance. Fig.\ref{Electrical}b is a bi-directional gate sweep of the conductivity, indicating low hysteresis and charge-trapping in the Au/Al$_2$O$_3$/SLG gate capacitor.
\begin{figure*}
\centerline{\includegraphics[width=180mm]{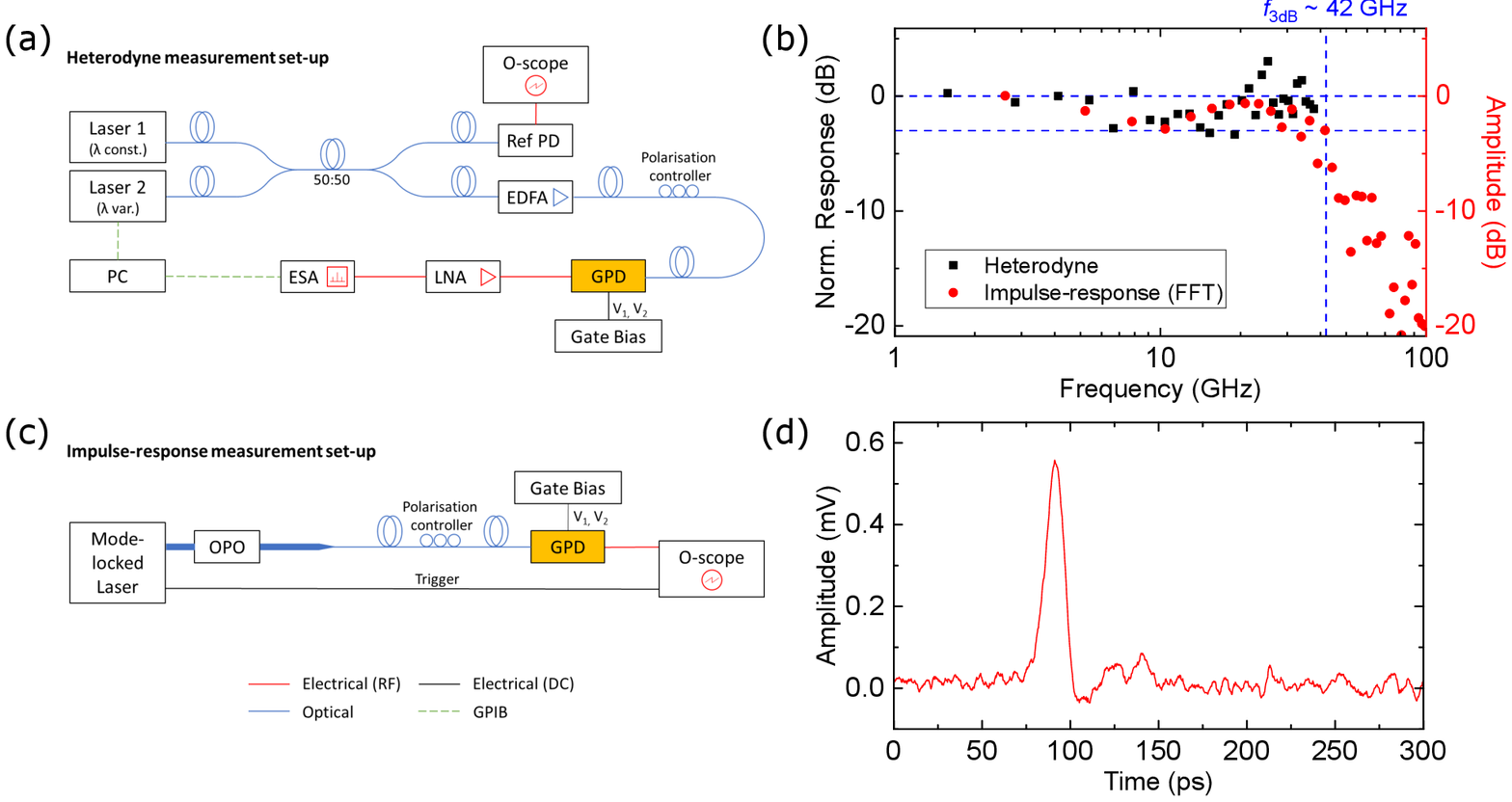}}
\caption{a) Schematic heterodyne set-up. b) Frequency response at zero $V_{DS}$ from (black) direct heterodyne measurement and (red) fast Fourier transform of impulse response. c) Schematic impulse-response measurement set-up. d) Impulse-response for$\sim$150fs optical pulses at zero $V_{DS}$ and p-n gate bias.}
\label{RF}
\end{figure*}

Fig.\ref{OpticalSgl}a is a $V_\text{ph}$ map of a typical device at $P_\text{in} \sim 100 \mu$W inside the GPD. The plot exhibits a six-fold pattern with higher response for bipolar (p-n, n-p) junctions and a weaker one with sign-crossing along the diagonal ($V_{\text{Gate 1}}=V_{\text{Gate 2}}$) for the unipolar (n-n, p-p) junctions. When the GPD is operated at zero $V_{DS}$, this indicates a PTE-dominated photodetection as the two sign changes in $V_\text{ph}$ along a single-gate sweep line (e.g $V_{\text{Gate 2}}=$const.) reflect the two sign changes of the $S$ gradient across the junction, arising from the non-monotonic dependence of $S$ on $\mu_c$\cite{Song2011Hot}. The measured photoresponse is in good agreement with the calculated one in Fig.\ref{OpticalSgl}b. We observed a similar behavior on$>12$ devices of different sizes across 5 chips, the shortest being 500nm in the light propagation direction for a footprint$\sim$3$\mu$m$^2$. For all devices, we got a maximum $V_\text{ph}$ close to the CNP where $S$ is largest, with a gradual drop-off at higher doping.

To calculate $R_V$, we first estimate the optical power inside our GPDs by taking into account: a) combined loss ($\sim$9.6dB at peak transmission) of waveguide propagation and fiber-to-waveguide coupling (wavelength-dependent, following the response envelope of the GC); b) 3dB power reduction from the input laser modulation (square wave, ON-OFF) with a 50\% duty cycle; c) 3dB power splitting in the Y-branches and their$\sim$0.2dB losses. We get $R_V\sim12.2$V/W, the largest reported so far for waveguide-integrated GPDs in voltage mode at zero-bias.

Fig.\ref{OpticalSgl}c plots the $R_V$ wavelength dependence, showing a broadband (1.50-1.58$\mu$m) photoresponse covering the entire C-band (1.53-1.565$\mu$m\cite{senior2009optical}) and beyond. The error bars indicate variations in the wavelength-dependent coupling loss (thus $P_\text{in}$), estimated as standard deviation from transmission measurements on$>$10 reference waveguides. We attribute the gradual increase in $R_V$ with increasing wavelengths to improved coupling efficiency from dielectric to plasmonic waveguide.

Fig.\ref{OpticalSgl}d is the $V_\text{ph}$ power dependence at 1.575$\mu$m for optical power levels comparable to those required by receivers used in 100GBs$^{-1}$ links\cite{Romagnoli2018Graphene}. The linear response indicates a power-independent $R_V$.

To highlight our GPDs' behavior as voltage sources, when a signal is generated, we place two devices back to back on the same waveguide and connect them in series. This modified design, Fig.\ref{OpticalDualConcept}a, consists of two SLG channels gated from the same split-gate/SPP waveguide. By connecting the drain pad of one channel to the source of the other through a metal lead crossing the waveguide behind the active region of the devices, Fig.\ref{OpticalDualConcept}b, we measure both GPDs individually, as well as combined. Fig.\ref{OpticalDualConcept}c is a false color SEM of the active region of both detectors. Since each GPD is designed to maximally absorb over the device length, the power rapidly decays along the propagation direction after the first GPD. We thus place the second device$\sim1\mu$m from the first.

Figs.\ref{OpticalDualConcept}d,e plot the photovoltage maps for both GPDs at $P_\text{in}\sim70\mu$W. The GPD closer to the input GC (A) absorbs most of the light and has the six-fold pattern typical of PTE, Fig.\ref{OpticalDualConcept}d. The photoresponse of the second GPD (B) is weaker, due to light absorption in SLG and metal, Fig.\ref{FDTD}d,e. A six-fold pattern is not observed, due to photocurrents generated in the junctions between gated and ungated sections at either end of the SLG channel. Figs.\ref{OpticalDualConcept}f,g are photovoltage maps of the combination of both GPDs. The response in series is in Fig.\ref{OpticalDualConcept}f, while the sum of the individual responses in Fig.\ref{OpticalDualConcept}g. The two plots are good agreement, confirming that $V_\text{ph,A+B}=V_\text{ph,A}+V_\text{ph,B}$. Thus, in order to increase $R_V$ in long (tens $\mu$m) PTE-GPDs with  absorption only in the SLG channel, one could avoid the reduction in photosignal due to the decrease in $\Delta T_e$ along the device and instead add the voltages generated in different sections by subdividing the channel into several shorter devices and connecting them as cascaded GPDs. To minimize the length of metal leads for contacting and connecting individual devices, this configuration could comprise individually gated devices with alternating p-n junctions to form a meandered structure. This would ideally be implemented with transparent gates, such as indium tin oxide, or a second SLG at a distance far enough from the channel, to avoid additional losses.

To evaluate the frequency response we use the optical heterodyne set-up in Fig.\ref{RF}a, combing optical signals at different frequencies. The channel is contacted with an RF probe in G-S configuration. The output of our tunable laser source is combined with that of a fixed-wavelength laser diode (Thorlabs SFL1550P) and the GPDs' response to the amplitude beating at the difference frequency is monitored with an electrical spectrum analyzer (ESA, Agilent PSX N9030A). The two source outputs are combined in a 50:50 fibre coupler. We monitor the signal stability (i.e. output power and the position of the difference frequency) using a reference PD and an oscilloscope. Prior to coupling the combined signal into the SiN waveguide, we use an erbium-doped fibre amplifier (EDFA, Keyopsys CEFA-C-HG) to raise the optical power to 15dBm ($\sim$30mW) to increase the output signal detected at the ESA. To overcome the signal reduction due to impedance mismatch between device $R$ and the 50$\Omega$ of the measurement equipment, we use an additional low noise amplifier (LNA) between GPD and ESA. In order to distinguish between the frequency response of our GPDs and that of the LNA and the remaining measurement equipment, the response of the latter two to a low power (-80dBm) input from a 50GHz signal generator is recorded for calibration.

Fig.\ref{RF}b plots the calibrated response (black squares) to the beating signal at different frequencies, while the split-gate is biased to set an operating point in the p-n junction regime resulting in the largest photoresponse under CW illumination in Fig.\ref{OpticalSgl}. The response stays within 3dB of the low-frequency (1GHz) reference power until 40GHz, the limit of our measurement set-up.

To determine the cut-off, we therefore modify the set-up (Fig.\ref{RF}c) to perform impulse response measurements, where the response to ultra-short ($\sim$150fs) optical pulses is monitored with an oscilloscope. For excitation, we use the idler of an optical parametric oscillator, pumped by a Ti:Sa mode-locked laser at 1.55$\mu$m, attenuated in free-space prior to coupling into the optical fibre. Fig.\ref{RF}d is the measured impulse response at the same operating point as for the heterodyne measurements. We obtain a pulse duration, assuming a Gaussian pulse shape, $\Delta t\sim11$ps. For Gaussian-shaped pulses, the time-bandwidth product is$\sim$0.44\cite{Diels2006Ultrashort}. From this we estimate a $f_\text{3dB}\sim0.44/\Delta t\sim40$GHz. The fast Fourier transform of the pulse is in Fig.\ref{RF}b (red circles) after calibration. The trace is in good agreement with the heterodyne response and drops below -3dB at$\sim42$GHz, showing high-speed operation on par with current Ge PDs, consistent with other reports of high-speed MC-SLG-based PTE GPDs\cite{Schuler2016Controlled,Shiue2015High,schuler2018graphene}. However, this bandwidth is the highest reported so far for PTE-based on-chip GPDs made from CVD SLG.

In summary, we reported waveguide-integrated plasmonic enhanced GPDs with an external responsivity$\sim$12.2V/W, a -3dB cut-off$\sim$42GHz, and small ($\sim$3-20$\mu$m$^2$) device footprints, using CVD SLG on SiN. We exploited the integration of an SPP waveguide with a SLG p-n junction to enhance light-SLG interaction and create a confined electron heat-source to obtain a strong, PTE-dominated photoresponse. This paves the way to power-efficient receivers for optoelectronic links.
\begin{acknowledgments}
We acknowledge funding from EU Graphene Flagship, ERC Grant Hetero2D, EPSRC Grants EP/K01711X/1, EP/K017144/1, EP/N010345/1, and EP/L016087/1.
\end{acknowledgments}
\section*{Methods}
\subsection{Device design and modeling}
Plasmonic gap width, SLG placement, and thickness of metal and oxide structures are the key parameters to be optimized to achieve maximum photovoltage. To do so, we build a device model in the layout environment of a commercial eigenmode solver (Lumerical MODE Solutions). In order to model SLG in the optical solver and subsequent calculations consistently, we use a volumetric permittivity material model, in which SLG is described as cuboid with finite thickness $t=0.34$nm and in-plane ($\varepsilon_\parallel$) and out-of plane ($\varepsilon_\perp$) permittivity are defined as independent tensor elements. To calculate the in-plane relative permittivity for SLG, we use\cite{emani2015graphene}:
\begin{equation} \label{eq:PermSurf}
\varepsilon(\omega) = \varepsilon'(\omega) + i \varepsilon''(\omega) = \varepsilon_r + \frac{i\sigma(\omega)}{\varepsilon_0\omega t}
\end{equation}
where $\varepsilon'(\omega)$ and $\varepsilon''(\omega)$ are the real and imaginary part of the relative permittivity $\varepsilon(\omega)$, $\sigma(\omega)$ is the SLG optical conductivity, $\omega$ is the angular frequency, $\varepsilon_r$ is the background relative permittivity (whose frequency-dependence is ignored in the small ($\lambda=$1.5-1.6$\mu$m) wavelength range under consideration), and $\varepsilon_0$ is the permittivity of free space. $\sigma(\omega)$ is obtained by linear-response\cite{hanson2008dyadic} in the random-phase approximation\cite{emani2015graphene}, and contains terms describing intraband and interband transitions. While the former can be evaluated analytically\cite{hanson2008dyadic}, the interband term requires a numerical solution\cite{hanson2008dyadic,emani2015graphene}. The out-of-plane component of the dielectric tensor matches $\varepsilon_r$.

We then run the eigenmode solver, using the relative permittivity function for SLG, and select the fundamental (gap plasmon) mode for further processing. We export the simulation mesh grid positions, electric $\mathbf{E}$ and magnetic $\mathbf{H}$  data, effective refractive index ($n_\text{eff}=\beta/k_0$, where $\beta$ is the propagation constant of the mode and $k_0$ is the free space wavevector\cite{reed2004silicon}), and all relevant geometric parameters such as gap and gate width and gate oxide thickness to complete the rest of our modeling.

A crucial intermediate step requires the determination of the $T_e(x)$ profile in the SLG channel. The first step establishes the operating regime. As discussed in Refs.\cite{Soavi2018Broadband,soavi2019hot}, the energy delivered to the electronic carrier distribution by pumping SLG with a pulsed laser can be sufficiently high to result in $T_e(x)>1000$K. When modeling photoexcited SLG under these conditions, one has to take into account the $T_e(x)$ dependence\cite{Soavi2018Broadband,soavi2019hot} of all thermodynamic and transport parameters in Eqs.\ref{eq:VPTE}-\ref{eq:heatequation}, i.e. $\mu_{c}$, $\sigma(x)$,$\kappa_{e}(x)$, resulting in a nonlinear system of coupled equations.
\begin{figure}[tb]
\centerline{\includegraphics[width=90mm]{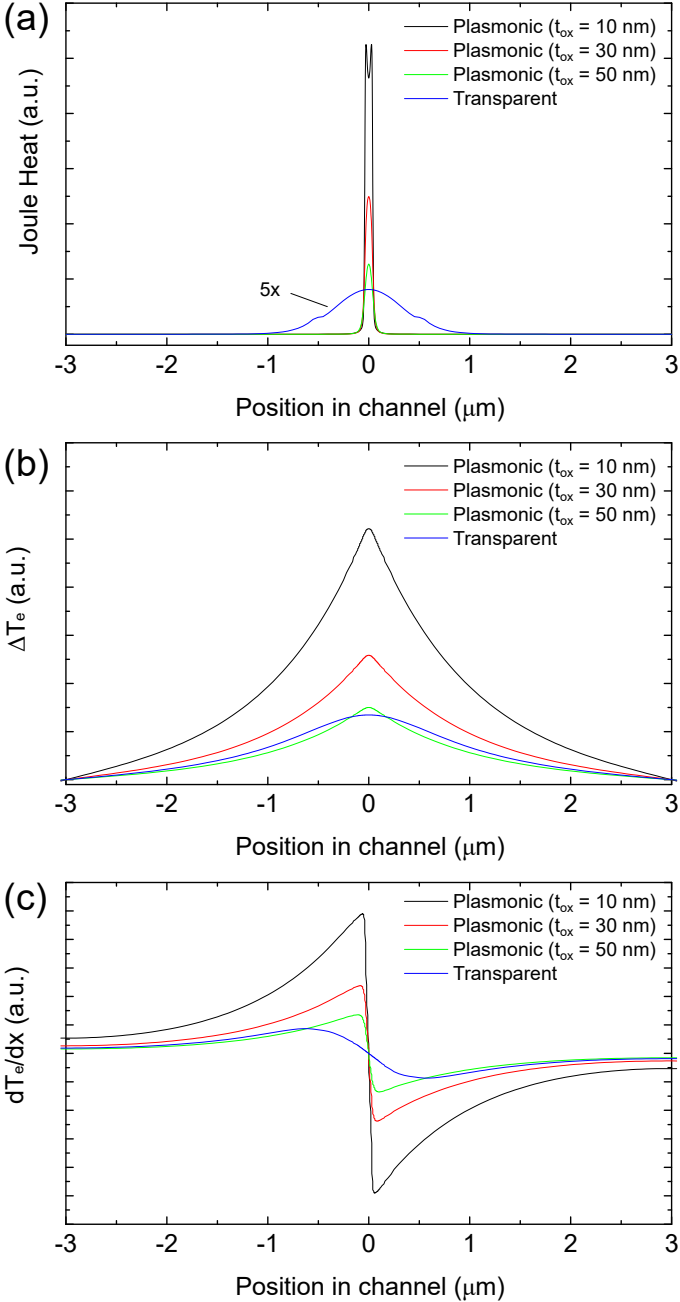}}
\caption{a) Joule heat source, b) $T_\text{e}$ profile, c) $\frac{\mathrm{d}T_e}{\mathrm{d}x}$ along SLG channel for a representative GPD with plasmonic gates at different heights above the channel compared to gates transparent at $1.55\mu$m}
\label{DesignSimPlot}
\end{figure}

The contrasting case, the ``weak heating'' regime, is characterized by $|\Delta T_e(x)|\ll T_l$\cite{tielrooij2015generation,castilla2019fast}. Under this condition, Eqs.\ref{eq:VPTE}-\ref{eq:heatequation} can be solved to linear order in the local $T_e(x)$ fluctuation, evaluating all thermodynamic and transport parameters at $T_{e}(x)=T_{l}$. In particular, a single $\mu_c$ is established, following photoexcitation, by electron-electron interactions across the valence and conduction bands, and the thermal conductivity, calculated from the Wiedemann-Franz law $\kappa_e=\pi^2 k_\text{B}^2 T_{e} \sigma/(3 e^2)$\cite{kittel1996introduction}, is uniform in space.

The intended operation of our GPD is under CW or, during data reception, quasi-CW (i.e. the pulse duration exceeds the cooling time of the hot-carrier distribution) illumination with low $P_\text{in}<0.1$mW at in-plane incidence. Furthermore, a linear dependence of $V_\text{ph}$ on $P_\text{in}$, such as in Fig.\ref{OpticalSgl}d, is observed\cite{tielrooij2015generation}. We thus conclude that our device operates in the ``weak heating'' regime.

In order to evaluate Eq.\ref{eq:heatequation}, we need to specify the cooling length $\xi$ and the light absorption heat source. In principle, $\xi$, which describes the energy transfer from the electronic system to the optical phonons\cite{Song2011Hot}, depends on $T_e(x)$ and $n$. However, under the ``weak heating'' regime, the $T_e(x)$ dependence can be neglected, and $n$ is uniform in the device (with opposite sign in the two regions of the p-n junction) when the photoresponse is maximal. For these reasons, in our calculations, we assume a constant $\xi=1\mu$m, consistent with experimental values\cite{Ma2014Competing,Shiue2015High}. $J(x)$ is calculated from the simulated electric field data as the time average of the electric power dissipation density\cite{desiatov2014direct,shin2012instantaneous}:	
\begin{equation}
J(x) = \frac{1}{2} \omega \epsilon'' \lvert \mathbf{E} \rvert^2 = \frac{1}{2} \frac{\sigma}{t}\mathbf{E}\cdot\mathbf{E^{\ast}}
\end{equation}  	
\begin{figure*}
\centerline{\includegraphics[width=180mm]{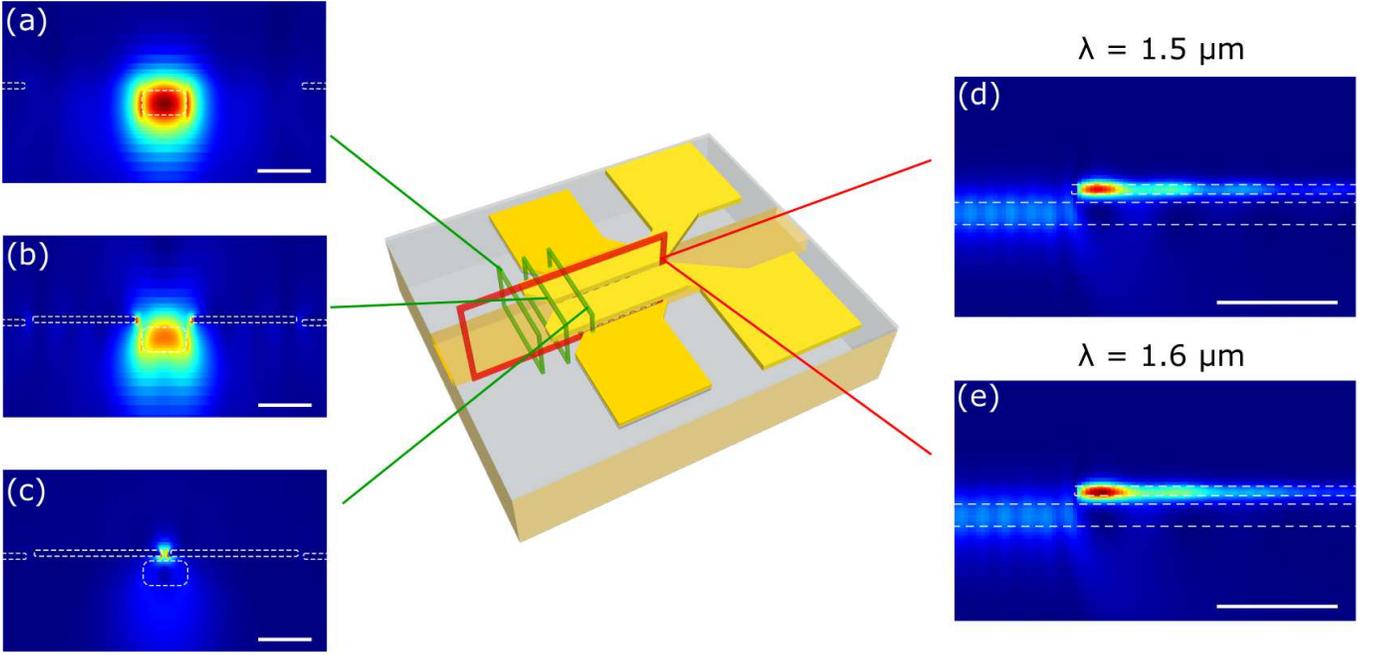}}
\caption{Field distribution a) before the GPD, b) during transition, and c) start of the split-gate/SPP waveguide. Scale bar: 1$\mu$m. (d,e) $\lvert\mathbf{E}\rvert^2$ along the center (red box) of dielectric waveguide and GPD at d)1.5$\mu$m and e) $=1.6\mu$m. Scale bar: 3$\mu$m}
\label{FDTD}
\end{figure*}
In order to relate $J(x)$ to physically meaningful quantities, we integrate the normal component of the time-averaged Poynting vector over the simulation region and normalize it to a given input power.

After solving Eq.\ref{eq:heatequation} for the $T_e(x)$ fluctuation profile $\Delta T_e(x)$, we take its derivative with respect to $x$ and obtain the second factor of the integrand in Eq.\ref{eq:VPTE}. Figs.\ref{DesignSimPlot}a-c compare the absorption heat source, the resulting $T_{e}(x)$ profile, and its derivative for a representative GPD in presence of a plasmonic split-gate at different heights over the SLG channel, to an unperturbed fundamental dielectric waveguide mode, where the p-n junction is generated by a transparent (at the chosen $\lambda$) gate, such as a split-gate made from a second SLG at a separation large enough to avoid additional optical losses. The beneficial role of plasmonic enhancement, with all other parameters fixed, for a sharper increase in $T_e(x)$ translates to larger $V_\text{ph}$ if the SLG channel is kept close to the SPP waveguide ($<$50nm).

To model $S$ along the channel, we assume that the structure is gated to achieve the maximum $S$ below the gates, as for Eq.\ref{eq:MottFormula} (opposite in sign but equal in magnitude for the p-n case) and approximate the gap region with a linear interpolation between the two. Combining both factors in Eq.\ref{eq:VPTE} and computing the integral yields $V_\text{ph}$ or $R_V$ (if divided by $P_\text{in}$) as figure of merit to assess different cross section designs:
\begin{equation} \label{eq:Rvfinal}
R_V = \frac{\lvert V_\text{ph} \rvert}{P_\text{in}} = \frac{1}{P_\text{in}} \bigg\lvert\int S(x) \frac{\mathrm{d}T_e(x)}{\mathrm{d}x}\; \mathrm{d} x \bigg\rvert
\end{equation}
To calculate $R_V$ or photovoltage maps as in Fig.\ref{OpticalSgl}b, we first generate the $S$ profile for all voltage combinations and then proceed via Eqs.\ref{eq:Rvfinal} or \ref{eq:VPTE}.

We then perform a sweep of gap width, SLG position, gate oxide thickness, gate contact height as a function of $\mu$ in SLG and $\varepsilon_r$. As for Fig.\ref{DesignSimPlot}, shorter distances between SLG channel and SPP waveguide yield larger signals. Furthermore, the SPP waveguide width affects the expected photovoltage in two ways. 1) narrower confinement improves the field strength at the SLG channel and $R_V$ at the cost of higher propagation losses; 2) the ungated SLG at the center of the device with not maximum $S$ shrinks with smaller gaps. The device parameters are then chosen based on these trends, taking into account fabrication complexity, robustness to processing-induced deviations (e.g. suppression of fundamental gap plasmon mode below a certain gap width), reliability (e.g. oxide break-down) and outcomes of FDTD simulations on coupling and propagation.

Extracting $\mathbf{E}$ and $\mathbf{H}$ of the fundamental mode in the hybrid region is sufficient for the design of the device cross-section. However, it does not capture the field distribution along the device, since optical losses, transition from dielectric to plasmonic waveguide, and power exchange between different modes that co-exist in the hybrid region remain unaccounted for. To find a good combination of SLG position and width within the constraints of the cross-section design, targeting a $T_e$ distribution with maximal derivative across the PD, but minimal along the propagation direction over the device length, we perform FDTD of the transition between SiN waveguide and GPD. We construct a device model in the same way as for the eigenmode analysis and adjust the source settings to 1.5-1.6$\mu$m. We launch the fundamental quasi-TE mode of the SiN waveguide towards the GPD and use frequency domain field monitors (FDFM) with various orientations (parallel and perpendicular to the propagation direction) to track the field and power profiles at different points. Figs.\ref{FDTD}a-c plot the field at 3 cross-sections of a representative GPD. We see transitions from injected mode in dielectric waveguide to field distribution resembling the fundamental gap plasmon.

Figs.\ref{FDTD}d,e display the electric field intensity from a vertical FDMD monitor along the center line of the GPD, at two wavelengths. As expected for this non-adiabatic transition with fast decrease ($<$1$\mu$m) of the taper cross-section down to the target gap size, scattering and reflections at the start of the hybrid region reduce the power at the GPD, but the desired sharp intensity profile over length scales that match the fabricated SLG channel widths is achieved. Consequently, we place the SLG channel 100nm after the SPP structure has reached its final gap width. The comparison of field intensities for 1.5 and 1.6$\mu$m on the same color scale in Figs.\ref{FDTD}d,e reveals a larger peak intensity and a longer interaction length for the latter, which indicates improved coupling efficiency at larger $\lambda$, as for the wavelength-depended $R_V$ in Fig.\ref{OpticalSgl}c.


\section*{References}
\begin{thebibliography}{100}
\bibitem{cisco2017visual}https://www.cisco.com/c/en/us/solutions/collateral/service-provider/visual-networking-index-vni/complete-white-paper-c11-481360.pdf
\bibitem{Andrews2014What}J. G. Andrews, S. Buzzi, W. Choi, S. V. Hanly, A. Lozano, A. C. K. Soong, and J. C. Zhang, IEEE J. Sel. Areas Commun. \textbf{32}, 1065 (2014).
\bibitem{Osseiran2014Scenarios}A. Osseiran, F. Boccardi, V. Braun, K. Kusume, P. Marsch, M. Maternia, O. Queseth, M. Schellmann, H. Schotten, H. Taoka \emph{et al.}, IEEE Commun. Mag. \textbf{52}, 26 (2014).
\bibitem{Arm2017route}https://community.arm.com/cfs-file/key/telligent-evolution-components-attachments/01-1996-00-00-00-01-30-09/Arm-2D00-The-route-to-a-trillion-devices-2D00-June-2017.pdf
\bibitem{Rumley2015Silicon}S. Rumley, D. Nikolova, R. Hendry, Q. Li, D. Calhoun, and K. Bergman, J. Light. Technol. \textbf{33}, 547 (2015).
\bibitem{zhou2018development}Z. Zhou, R. Chen, X. Li, and T. Li, Opt. Fiber Technol. \textbf{44}, 13 (2018).
\bibitem{reed2008silicon}G. T. Reed, \emph{Silicon Photonics: The State of the Art} (John Wiley \& Sons, Chichester, 2008).
\bibitem{thomson2016roadmap}D. Thomson, A. Zilkie, J. E. Bowers, T. Komljenovic, G. T. Reed, L. Vivien, D. Marris-Morini, E. Cassan, L. Virot, J. F{\'e}d{\'e}li \emph{et al.}, J. Opt. \textbf{18}, 073003 (2016).
\bibitem{absil2015silicon}P. P. Absil, P. Verheyen, P. De Heyn, M. Pantouvaki, G. Lepage, J. De Coster, and J. Van Campenhout, Opt. Express \textbf{23}, 9369 (2015).
\bibitem{Atabaki2018Integrating}A. H. Atabaki, S. Moazeni, F. Pavanello, H. Gevorgyan, J. Notaros, L. Alloatti, M. T. Wade, C. Sun, S. A. Kruger, H. Meng \emph{et al.}, Nature \textbf{556}, 349 (2018).
\bibitem{biberman2012optical}A. Biberman, and K. Bergman, Rep. Prog. Phys. \textbf{75}, 046402 (2012).
\bibitem{Wooten2000review}E. L. Wooten, K. M. Kissa, A. Yi-Yan, E. J. Murphy, D. A. Lafaw, P. F. Hallemeier, D. Maack, D. V. Attanasio, D. J. Fritz, G. J. McBrien \emph{et al.}, IEEE J. Sel. Topics Quantum Electron. \textbf{6}, 69 (2000).
\bibitem{Nagarajan2010InP}R. Nagarajan, M. Kato, J. Pleumeekers, P. Evans, S. Corzine, S. Hurtt, A. Dentai, S. Murthy, M. Missey, R. Muthiah \emph{et al.}, IEEE J. Sel. Topics Quantum Electron. \textbf{16}, 1113 (2010).
\bibitem{Reed2010silicon}G. T. Reed, G. Mashanovich, F. Y. Gardes, and D. J. Thomson, Nat. Photonics \textbf{4}, 518 (2010).
\bibitem{reed2014recent}G. T. Reed, G. Z. Mashanovich, F. Y. Gardes, M. Nedeljkovic, Y. Hu, D. J. Thomson, K. Li, P. R. Wilson, S. Chen, and S. S. Hsu, Nanophotonics \textbf{3}, 229 (2014).
\bibitem{seraphin1965franz}B. Seraphin, and N. Bottka, Phys. Rev. \textbf{139}, A560 (1965).
\bibitem{liu2008waveguide}J. Liu, M. Beals, A. Pomerene, S. Bernardis, R. Sun, J. Cheng, L. C. Kimerling, and J. Michel, Nat. Photonics \textbf{2}, 433 (2008).
\bibitem{srinivasan201656}S. A. Srinivasan, M. Pantouvaki, S. Gupta, H. T. Chen, P. Verheyen, G. Lepage, G. Roelkens, K. Saraswat, D. Van Thourhout, P. Absil \emph{et al.}, J. Light. Technol. \textbf{34}, 419 (2016).
\bibitem{Michel2010High}J. Michel, J. Liu, and L. C. Kimerling, Nat. Photonics \textbf{4}, 527 (2010).
\bibitem{Hawkins1997High}A. R. Hawkins, W. Wu, P. Abraham, K. Streubel, and J. E. Bowers, Appl. Phys. Lett. \textbf{70}, 303 (1997).
\bibitem{chang2010integrated}H. Chang, Y. Kuo, R. Jones, A. Barkai, and J. E. Bowers, Opt. Express \textbf{18}, 23891 (2010).
\bibitem{chrostowski2015silicon}L. Chrostowski, and M. Hochberg, \emph{Silicon Photonics Design: From Devices to Systems} (Cambridge University Press, Glasgow, 2015).
\bibitem{vivien2009Ghz}L. Vivien, J. Osmond, J. F{\'e}d{\'e}li, D. Marris-Morini, P. Crozat, J. Damlencourt, E. Cassan, Y. Lecunff, and S. Laval, Opt. Express \textbf{17}, 6252 (2009).
\bibitem{derose2011ultra}C. T. DeRose, D. C. Trotter, W. A. Zortman, A. L. Starbuck, M. Fisher, M. R. Watts, and P. S. Davids, Opt. Express \textbf{19}, 24897 (2011).
\bibitem{vivien2012zero}L. Vivien, A. Polzer, D. Marris-Morini, J. Osmond, J. M. Hartmann, P. Crozat, E. Cassan, C. Kopp, H. Zimmermann, and J. M. F{\'e}d{\'e}li, Opt. Express \textbf{20}, 1096 (2012).
\bibitem{novack2013germanium}A. Novack, M. Gould, Y. Yang, Z. Xuan, M. Streshinsky, Y. Liu, G. Capellini, A. E. Lim, G. Lo, T. Baehr-Jones \emph{et al.}, Opt. Express \textbf{21}, 28387 (2013).
\bibitem{Chen2016Bias}H. Chen, P. Verheyen, P. De Heyn, G. Lepage, J. De Coster, S. Balakrishnan, W. Yao, L. Shen, G. Roelkens, and J. Van Campenhout, Opt. Express \textbf{24}, 4622 (2016).
\bibitem{wang2011ge}J. Wang, and S. Lee, Sensors \textbf{11}, 696 (2011).
\bibitem{ye2014germanium}H. Ye, and J. Yu, Sci. Technol. Adv. Mater. \textbf{15}, 024601 (2014).
\bibitem{sorianello2012high}V. Sorianello, A. De Iacovo, L. Colace, A. Fabbri, L. Tortora, E. Buffagni, and G. Assanto, Appl. Phys. Lett. \textbf{101}, 081101 (2012).
\bibitem{liu2005tensile}J. Liu, D. D. Cannon, K. Wada, Y. Ishikawa, S. Jongthammanurak, D. T. Danielson, J. Michel, and L. C. Kimerling, Appl. Phys. Lett. \textbf{87}, 011110 (2005).
\bibitem{Romagnoli2018Graphene}M. Romagnoli, V. Sorianello, M. Midrio, F. H. L. Koppens, C. Huyghebaert, D. Neumaier, P. Galli, W. Templ, A. D'Errico, and A. C. Ferrari, Nat. Rev. Mater. \textbf{3}, 392–414 (2018).
\bibitem{Liu2011graphene}M. Liu, X. Yin, E. Ulin-Avila, B. Geng, T. Zentgraf, L. Ju, F. Wang, and X. Zhang, Nature \textbf{474}, 64 (2011).
\bibitem{Liu2012double}M. Liu, X. Yin, and X. Zhang, Nano Lett. \textbf{12}, 1482 (2012).
\bibitem{Hu2016broadband}Y. Hu, M. Pantouvaki, J. {Van Campenhout}, S. Brems, I. Asselberghs, C. Huyghebaert, P. Absil, and D. {Van Thourhout}, Laser Photon. Rev. \textbf{10}, 307 (2016).
\bibitem{phare2015graphene}C. T. Phare, Y. Daniel Lee, J. Cardenas, and M. Lipson, Nat. Photonics \textbf{9}, 511 (2015).
\bibitem{sorianello2015design}V. Sorianello, M. Midrio, and M. Romagnoli, Opt. Express \textbf{23}, 6478 (2015).
\bibitem{Sorianello2018graphene}V. Sorianello, M. Midrio, G. Contestabile, I. Asselberghs, J. Van Campenhout, C. Huyghebaert, I. Goykhman, A. K. Ott, A. C. Ferrari, and M. Romagnoli, Nat. Photonics \textbf{12}, 40 (2018).
\bibitem{Sun2016optical}Z. Sun, A. Martinez, and F. Wang, Nat. Photonics \textbf{10}, 227 (2016).
\bibitem{Gan2013chip}X. Gan, R. Shiue, Y. Gao, I. Meric, T. F. Heinz, K. Shepard, J. Hone, S. Assefa, and D. Englund, Nat. Photonics \textbf{7}, 883 (2013).
\bibitem{Pospischil2013CMOS}A. Pospischil, M. Humer, M. M. Furchi, D. Bachmann, R. Guider, T. Fromherz, and T. Mueller, Nat. Photonics \textbf{7}, 892 (2013).
\bibitem{Wang2013high}X. Wang, Z. Cheng, K. Xu, H. K. Tsang, and J. Xu, Nat. Photonics \textbf{7}, 888 (2013).
\bibitem{Goykhman2016On}I. Goykhman, U. Sassi, B. Desiatov, N. Mazurski, S. Milana, D. {De Fazio}, A. Eiden, J. Khurgin, J. Shappir, U. Levy \emph{et al.}, Nano Lett. \textbf{16}, 3005 (2016).
\bibitem{Schall2014GBits}D. Schall, D. Neumaier, M. Mohsin, B. Chmielak, J. Bolten, C. Porschatis, A. Prinzen, C. Matheisen, W. Kuebart, B. Junginger \emph{et al.}, ACS Photonics \textbf{1}, 781 (2014).
\bibitem{Schuler2016Controlled}S. Schuler, D. Schall, D. Neumaier, L. Dobusch, O. Bethge, B. Schwarz, M. Krall, and T. Mueller, Nano Lett. \textbf{16}, 7107 (2016).
\bibitem{schuler2018graphene}S. Schuler, D. Schall, D. Neumaier, B. Schwarz, K. Watanabe, T. Taniguchi, and T. Mueller, ACS Photonics \textbf{5}, 4758 (2018).
\bibitem{Shiue2015High}R. J. Shiue, Y. Gao, Y. Wang, C. Peng, A. D. Robertson, D. K. Efetov, S. Assefa, F. H. L. Koppens, J. Hone, and D. Englund, Nano Lett. \textbf{15}, 7288 (2015).
\bibitem{Schall2017graphene}D. Schall, C. Porschatis, M. Otto, and D. Neumaier, J. Phys. D: Appl. Phys. \textbf{50}, 124004 (2017).
\bibitem{Schall2018record}D. Schall, E. Pallecchi, G. Ducournau, V. Avramovic, M. Otto, and D. Neumaier, in \emph{Optical Fiber Communication Conference}, M2I.4, (Optical Society of America, 2018).
\bibitem{Ding2018ultra}Y. Ding, Z. Cheng, X. Zhu, K. Yvind, J. Dong, M. Galili, H. Hu, N. A. Mortensen, S. Xiao, and L. K. Oxenl{\o}we, \emph{arXiv}: 1808.04815 (2018).
\bibitem{Ma2018Plasmonically}P. Ma, Y. Salamin, B. Baeuerle, A. Josten, W. Heni, A. Emboras, and J. Leuthold, ACS Photonics \textbf{6}, 154 (2019).
\bibitem{ma2018compact}Z. Ma, K. Kikunage, H. Wang, S. Sun, R. Amin, M. Tahersima, R. Maiti, M. Miscuglio, H. Dalir, and V. J. Sorger, \emph{arXiv}: 1812.00894 (2018).
\bibitem{Bonaccorso2010Graphene}F. Bonaccorso, Z. Sun, T. Hasan, and A. C. Ferrari, Nat. Photonics \textbf{4}, 611 (2010).
\bibitem{urich2011intrinsic}A. Urich, K. Unterrainer, and T. Mueller, Nano Lett. \textbf{11}, 2804 (2011).
\bibitem{Xia2009ultrafast}F. Xia, T. Mueller, Y. Lin, A. Valdes-Garcia, and P. Avouris, Nat. Nanotechnol. \textbf{4}, 839 (2009).
\bibitem{Nair2008Fine}R. R. Nair, P. Blake, A. N. Grigorenko, K. S. Novoselov, T. J. Booth, T. Stauber, N. M. R. Peres, and A. K. Geim, Science \textbf{320}, 1308 (2008).
\bibitem{dawlaty2008measurement}J. M. Dawlaty, S. Shivaraman, J. Strait, P. George, M. Chandrashekhar, F. Rana, M. G. Spencer, D. Veksler, and Y. Chen, Appl. Phys. Lett. \textbf{93}, 131905 (2008).
\bibitem{Koppens2014Photodetectors}F. H. Koppens, T. Mueller, P. Avouris, A. C. Ferrari, M. S. Vitiello, and M. Polini, Nat. Nanotechnol. \textbf{9}, 780 (2014).
\bibitem{goossens2017broadband}S. Goossens, G. Navickaite, C. Monasterio, S. Gupta, J. J. Piqueras, R. P{\'e}rez, G. Burwell, I. Nikitskiy, T. Lasanta, T. Gal{\'a}n \emph{et al.}, Nat. Photonics \textbf{11}, 366 (2017).
\bibitem{Youngblood2016integration}N. Youngblood, and M. Li, Nanophotonics \textbf{6}, 1205 (2017).
\bibitem{Liu2013silicon}M. Liu, and X. Zhang, Nat. Photonics \textbf{7}, 851 (2013).
\bibitem{tielrooij2015generation}K. Tielrooij, L. Piatkowski, M. Massicotte, A. Woessner, Q. Ma, Y. Lee, K. S. Myhro, C. N. Lau, P. Jarillo-Herrero, N. F. van Hulst \emph{et al.}, Nat. Nanotechnol. \textbf{10}, 437 (2015).
\bibitem{Mueller2010graphene}T. Mueller, F. Xia, and P. Avouris, Nat. Photonics \textbf{4}, 297 (2010).
\bibitem{Echtermeyer2014Photothermoelectric}T. J. Echtermeyer, P. S. Nene, M. Trushin, R. V. Gorbachev, A. L. Eiden, S. Milana, Z. Sun, J. Schliemann, E. Lidorikis, K. S. Novoselov \emph{et al.}, Nano Lett. \textbf{14}, 3733 (2014).
\bibitem{Song2011Hot}J. C. W. Song, M. S. Rudner, C. M. Marcus, and L. S. Levitov, Nano Lett. \textbf{11}, 4688 (2011).
\bibitem{gabor2011hot}N. M. Gabor, J. C. Song, Q. Ma, N. L. Nair, T. Taychatanapat, K. Watanabe, T. Taniguchi, L. S. Levitov, and P. Jarillo-Herrero, Science \textbf{334}, 648 (2011).
\bibitem{Konstantatos2012hybrid}G. Konstantatos, M. Badioli, L. Gaudreau, J. Osmond, M. Bernechea, F. P. G. de Arquer, F. Gatti, and F. H. L. Koppens, Nat. Nanotechnol. \textbf{7}, 363 (2012).
\bibitem{Vicarelli2012graphene}L. Vicarelli, M. S. Vitiello, D. Coquillat, A. Lombardo, A. C. Ferrari, W. Knap, M. Polini, V. Pellegrini, and A. Tredicucci, Nat. Mater. \textbf{11}, 865 (2012).
\bibitem{freitag2013photoconductivity}M. Freitag, T. Low, F. Xia, and P. Avouris, Nat. Photonics \textbf{7}, 53 (2013).
\bibitem{Sassi2016graphene}U. Sassi, R. Parret, S. Nanot, M. Bruna, S. Borini, D. {De Fazio}, Z. Zhao, E. Lidorikis, F. H. L. Koppens, A. C. Ferrari \emph{et al.}, Nat. Commun. \textbf{8}, 14311 (2017).
\bibitem{Huo2018Recent}N. Huo, and G. Konstantatos, Adv. Mater. \textbf{30}, 1801164 (2018).
\bibitem{Brida2013ultrafast}D. Brida, A. Tomadin, C. Manzoni, Y. J. Kim, A. Lombardo, S. Milana, R. R. Nair, K. S. Novoselov, A. C. Ferrari, G. Cerullo \emph{et al.}, Nat. Commun. \textbf{4}, 1987 (2013).
\bibitem{tomadin2013nonequilibrium}A. Tomadin, D. Brida, G. Cerullo, A. C. Ferrari, and M. Polini, Phys. Rev. B \textbf{88}, 035430 (2013).
\bibitem{kittel1996introduction}C. Kittel, \emph{Introduction to solid state physics} (Wiley New York, New York, 1996).
\bibitem{bonini2007phonon}N. Bonini, M. Lazzeri, N. Marzari, and F. Mauri, Phys. Rev. Lett. \textbf{99}, 176802 (2007).
\bibitem{lazzeri2005electron}M. Lazzeri, S. Piscanec, F. Mauri, A. Ferrari, and J. Robertson, Phys. Rev. Lett. \textbf{95}, 236802 (2005).
\bibitem{ashcroft1976solid}N. Ashcroft, and N. Mermin, \emph{Solid State Physics} (Harcourt College Publishers, Fort Worth, 1976).
\bibitem{Ma2014Competing}Q. Ma, N. M. Gabor, T. I. Andersen, N. L. Nair, K. Watanabe, T. Taniguchi, and P. Jarillo-Herrero, Phys. Rev. Lett. \textbf{112}, 247401 (2014).
\bibitem{freitag2013increased}M. Freitag, T. Low, and P. Avouris, Nano Lett. \textbf{13}, 1644 (2013).
\bibitem{Herring2014photoresponse}P. K. Herring, A. L. Hsu, N. M. Gabor, Y. C. Shin, J. Kong, T. Palacios, and P. Jarillo-Herrero, Nano Lett. \textbf{14}, 901 (2014).
\bibitem{castilla2019fast}S. Castilla, B. Terres, M. Autore, L. Viti, J. Li, A. Nikitin, I. Vangelidis, K. Watanabe, T. Taniguchi, E. Lidorikis \emph{et al.}, Nano Lett. \textbf{},  (2019).
\bibitem{Novoselov2005Two}K. S. Novoselov, D. Jiang, F. Schedin, T. J. Booth, V. V. Khotkevich, S. V. Morozov, and A. K. Geim, Proc. Natl. Acad. Sci. U.S.A. \textbf{102}, 10451 (2005).
\bibitem{Wang2013One}L. Wang, I. Meric, P. Y. Huang, Q. Gao, Y. Gao, H. Tran, T. Taniguchi, K. Watanabe, L. M. Campos, D. A. Muller \emph{et al.}, Science \textbf{342}, 614 (2013).
\bibitem{purdie2018cleaning}D. G. Purdie, N. M. Pugno, T. Taniguchi, K. Watanabe, A. C. Ferrari, and A. Lombardo, Nat. Commun. \textbf{9}, 5387 (2018).
\bibitem{DeFazio2019high}D. De Fazio, D. G. Purdie, A. K. Ott, P. Braeuninger-Weimer, T. Khodkov, S. Goossens, T. Taniguchi, K. Watanabe, P. Livreri, F. H. L. Koppens \emph{et al.}, \emph{arXiv}: 1904.01405 (2019).
\bibitem{Miseikis2017deterministic}V. Miseikis, F. Bianco, J. David, M. Gemmi, V. Pellegrini, M. Romagnoli, and C. Coletti, 2D Mater. \textbf{4}, 021004 (2017).
\bibitem{li2010graphene}X. Li, C. W. Magnuson, A. Venugopal, J. An, J. W. Suk, B. Han, M. Borysiak, W. Cai, A. Velamakanni, Y. Zhu \emph{et al.}, Nano Lett. \textbf{10}, 4328 (2010).
\bibitem{Wang2016Support}B. Wang, M. Huang, L. Tao, S. H. Lee, A. Jang, B. Li, H. S. Shin, D. Akinwande, and R. S. Ruoff, ACS Nano \textbf{10}, 1404 (2016).
\bibitem{suk2011transfer}J. W. Suk, A. Kitt, C. W. Magnuson, Y. Hao, S. Ahmed, J. An, A. K. Swan, B. B. Goldberg, and R. S. Ruoff, ACS Nano \textbf{5}, 6916 (2011).
\bibitem{Goykhman2011Locally}I. Goykhman, B. Desiatov, J. Khurgin, J. Shappir, and U. Levy, Nano Lett. \textbf{11}, 2219 (2011).
\bibitem{Goykhman2012Waveguide}I. Goykhman, B. Desiatov, J. Khurgin, J. Shappir, and U. Levy, Opt. Express \textbf{20}, 28594 (2012).
\bibitem{Salamin2018GHz}Y. Salamin, P. Ma, B. Baeuerle, A. Emboras, Y. Fedoryshyn, W. Heni, B. Cheng, A. Josten, and J. Leuthold, ACS Photonics \textbf{5}, 3291 (2018).
\bibitem{Echtermeyer2011Strong}T. J. Echtermeyer, L. Britnell, P. K. Jasnos, A. Lombardo, R. V. Gorbachev, A. N. Grigorenko, A. K. Geim, A. C. Ferrari, and K. S. Novoselov, Nat. Commun. \textbf{2}, 455 (2011).
\bibitem{fang2012graphene}Z. Fang, Z. Liu, Y. Wang, P. M. Ajayan, P. Nordlander, and N. J. Halas, Nano Lett. \textbf{12}, 3808 (2012).
\bibitem{Echtermeyer2016Surface}T. J. Echtermeyer, S. Milana, U. Sassi, A. Eiden, M. Wu, E. Lidorikis, and A. C. Ferrari, Nano Lett. \textbf{16}, 8 (2016).
\bibitem{chen2017three}C. Chen, N. Youngblood, R. Peng, D. Yoo, D. A. Mohr, T. W. Johnson, S. Oh, and M. Li, Nano Lett. \textbf{17}, 985 (2017).
\bibitem{desiatov2014direct}B. Desiatov, I. Goykhman, and U. Levy, Nano Lett. \textbf{14}, 648 (2014).
\bibitem{shin2012instantaneous}W. Shin, A. Raman, and S. Fan, J. Opt. Soc. Am. B \textbf{29}, 1048 (2012).
\bibitem{tian2009broadband}J. Tian, S. Yu, W. Yan, and M. Qiu, Appl. Phys. Lett. \textbf{95}, 013504 (2009).
\bibitem{dabos2018water}G. Dabos, D. Ketzaki, A. Manolis, E. Chatzianagnostou, L. Markey, J. Weeber, A. Dereux, A. L. Giesecke, C. Porschatis, B. Chmielak \emph{et al.}, IEEE Photon. J. \textbf{10}, 2700308 (2018).
\bibitem{li2010structurally}Q. Li, and M. Qiu, Opt. Express \textbf{18}, 15531 (2010).
\bibitem{delacour2010efficient}C. Delacour, S. Blaize, P. Grosse, J. M. Fedeli, A. Bruyant, R. Salas-Montiel, G. Lerondel, and A. Chelnokov, Nano Lett. \textbf{10}, 2922 (2010).
\bibitem{lagatsky20132}A. Lagatsky, Z. Sun, T. Kulmala, R. Sundaram, S. Milana, F. Torrisi, O. Antipov, Y. Lee, J. Ahn, C. Brown \emph{et al.}, Appl. Phys. Lett. \textbf{102}, 013113 (2013).
\bibitem{Ferrari2006Raman}A. C. Ferrari, J. C. Meyer, V. Scardaci, C. Casiraghi, M. Lazzeri, F. Mauri, S. Piscanec, D. Jiang, K. S. Novoselov, S. Roth \emph{et al.}, Phys. Rev. Lett. \textbf{97}, 187401 (2006).
\bibitem{Ferrari2013Raman}A. C. Ferrari, and D. M. Basko, Nat. Nanotechnol. \textbf{8}, 235 (2013).
\bibitem{Das2008Monitoring}A. Das, S. Pisana, B. Chakraborty, S. Piscanec, S. K. Saha, U. V. Waghmare, K. S. Novoselov, H. R. Krishnamurthy, A. K. Geim, A. C. Ferrari \emph{et al.}, Nat. Nanotechnol. \textbf{3}, 210 (2008).
\bibitem{Basko2009Electron}D. M. Basko, S. Piscanec, and A. C. Ferrari, Phys. Rev. B \textbf{80}, 165413 (2009).
\bibitem{Novoselov2004Electric}K. S. Novoselov, A. K. Geim, S. V. Morozov, D. Jiang, Y. Zhang, S. V. Dubonos, I. V. Grigorieva, and A. A. Firsov, Science \textbf{306}, 666 (2004).
\bibitem{senior2009optical}J. M. Senior, and M. Y. Jamro, \emph{Optical fiber communications: principles and practice} (Pearson Education, Harlow, 2009).
\bibitem{Diels2006Ultrashort}J. Diels, and W. Rudolph, \emph{Ultrashort laser pulse phenomena} (Elsevier, London, 2006).
\bibitem{emani2015graphene}N. K. Emani, A. V. Kildishev, V. M. Shalaev, and A. Boltasseva, Nanophotonics \textbf{4}, 214 (2015).
\bibitem{hanson2008dyadic}G. W. Hanson, J. Appl. Phys. \textbf{103}, 064302 (2008); {\it ibid.} \textbf{113}, 029902 (2013).
\bibitem{reed2004silicon}G. T. Reed, and A. P. Knights, \emph{Silicon Photonics: An Introduction} (John Wiley \& Sons, Chichester, 2004).
\bibitem{Soavi2018Broadband}G. Soavi, G. Wang, H. Rostami, D. G. Purdie, D. De Fazio, T. Ma, B. Luo, J. Wang, A. K. Ott, D. Yoon \emph{et al.}, Nat. Nanotechnol. \textbf{13}, 583 (2018).
\bibitem{soavi2019hot}G. Soavi, G. Wang, H. Rostami, A. Tomadin, O. Balci, I. Paradeisanos, E. Pogna, G. Cerullo, E. Lidorikis, M. Polini \emph{et al.}, \emph{arXiv}: 1903.00989 (2019).
\end{thebibliography}
\end{document}